\documentclass{ieeeaccess}
\usepackage{newtxtext}
\usepackage{cite}
\usepackage{amsmath,amssymb,amsfonts}
\usepackage{graphicx}
\usepackage{textcomp}
\usepackage{mathtools} 
\usepackage{algorithmic}
\usepackage{algorithm}
\usepackage{pdfcomment} 
\usepackage[dvipsnames]{xcolor}
\usepackage{soul}
\usepackage{tabularx}
\usepackage{balance}

\usepackage{bm}
\usepackage[caption=false,font=footnotesize,justification=centering]{subfig}
\makeatletter
\AtBeginDocument{\DeclareMathVersion{bold}
\SetSymbolFont{operators}{bold}{T1}{times}{b}{n}
\SetSymbolFont{NewLetters}{bold}{T1}{times}{b}{it}
\SetMathAlphabet{\mathrm}{bold}{T1}{times}{b}{n}
\SetMathAlphabet{\mathit}{bold}{T1}{times}{b}{it}
\SetMathAlphabet{\mathbf}{bold}{T1}{times}{b}{n}
\SetMathAlphabet{\mathtt}{bold}{OT1}{pcr}{b}{n}
\SetSymbolFont{symbols}{bold}{OMS}{cmsy}{b}{n}
\renewcommand\boldmath{\@nomath\boldmath\mathversion{bold}}}
\makeatother

\def\BibTeX{{\rm B\kern-.05em{\sc i\kern-.025em b}\kern-.08em
    T\kern-.1667em\lower.7ex\hbox{E}\kern-.125emX}}

\begin{document}
\history{Date of publication xxxx 00, 0000, date of current version xxxx 00, 0000.}
\doi{10.1109/ACCESS.2024.0429000}

\title{Weighted Flow Matching and Physics-Informed Nonlinear Filtering for Parameter Estimation in Digital Twins}
\author{\uppercase{Yasar Yanik}\authorrefmark{1},
\uppercase{Himadri Basu}\authorrefmark{2},
\uppercase{Ricardo G. Sanfelice}\authorrefmark{2},
and \uppercase{Daniele Venturi}\authorrefmark{1}}

\address[1]{Department of Applied Mathematics, University of California, Santa Cruz, CA 95064, USA}
\address[2]{Department of Electrical and Computer Engineering, University of California, Santa Cruz, CA 95064, USA}
\tfootnote{This research was supported by the U.S. Air Force Office of Scientific Research (AFOSR) SURI  ``Digital Twin Enabled Autonomous Control for On-Orbit Spacecraft Servicing'', contract number FA9550-23-1-0678}

\markboth
{Yanik \MakeLowercase{\textit{et al.}}: Weighted Flow Matching and Physics-Informed Nonlinear Filtering for Parameter Estimation in Digital Twins}
{Yanik \MakeLowercase{\textit{et al.}}: Weighted Flow Matching and Physics-Informed Nonlinear Filtering for Parameter Estimation in Digital Twins}

\corresp{Corresponding author: Daniele Venturi (e-mail: venturi@ucsc.edu).}

\begin{abstract}
Digital twins (DTs) rely on continuous synchronization between physical systems and their virtual counterparts through online parameter estimation under uncertainty. In many practical settings, however, this task is challenged by low observability, weak excitation, nonlinear dynamics, and noisy or biased measurements. In this work, we develop a new mathematical framework that integrates Weighted Flow Matching (WFM) generative modeling with physics-informed nonlinear filtering to enhance parameter estimation in DTs. WFM relies on dynamic reweighting of training samples, which guides the generative model toward parameter regimes most informative of the evolving system state.
This generative component is tightly coupled with a physics-informed filtering architecture based on the Unscented Kalman Filter (UKF), yielding a unified DT framework that combines data-driven probability transport with physically consistent state and parameter estimation.
The effectiveness of the new integrated framework is demonstrated within a spacecraft DT architecture, where stable moment of inertia estimation is achieved under uncertain and noisy sensing, with significant performance improvements over established approaches such as Extended Kalman Filtering (EKF) and Ensemble Kalman Filtering (EnKF). These results highlight the potential of weighted generative modeling as a core mechanism for real-time DT synchronization in operational and mission-critical systems.
\end{abstract}


\titlepgskip=-21pt

\maketitle
\section{Introduction}
\label{sec:introduction}
A digital twin (DT) is a virtual counterpart of a physical system/asset, enabling two-way communication and continuous co-evolution through reciprocal interactions \cite{yanik2024verification, yanik2024applying, yuce2022prognostics}. Rather than conceiving the DT as a fixed mirror of the physical system, a predictive DT is designed as an evolving framework that combines forecasting models, online calibration/synchronization, data assimilation, and control routines. 
As measurements are acquired from the physical asset, they are used to update model parameters together with their associated uncertainties \cite{yanik2019quantification, yanik2018uncertainty}. This feedback loop allows the DT to continuously refine its representation of the system state, maintaining synchronization with the true dynamics and enabling reliable decision-making \cite{henao2025digital, fernandes2024bwb}.

The main goal of this paper is to develop a new mathematical framework to enhance this synchronization capability, by leveraging generative modeling algorithms as a means to improve parameter estimation, uncertainty quantification, and overall predictive robustness in DTs. In particular, we integrate a new version of flow matching (FM) \cite{lipman2022flow}, which we call Weighted Flow Matching (WFM), with nonlinear filtering techniques to achieve robust parameter estimation and improved coherence between the physical and virtual systems, thereby turning generative modeling into a core mechanism for real-time DT synchronization rather than just an auxiliary inference component. 

Central to this new viewpoint is the use of differential equations to drive a continuous transformation between probability densities, mapping samples from an initial distribution $\rho_0$ toward a target distribution $\rho_1$, or vice versa \cite{ho2020denoising}. 
Flow Matching learns such probability paths by regressing onto vector fields that connect distributions through the minimization of suitable per-example objectives. This generalizes classical diffusion models  \cite{ho2020denoising,song2020score,Tatsuoka2025} and includes optimal transport interpolants, where straight-line flows usually yield faster, more stable training and improved generative performance compared to diffusion-based paths \cite{lipman2022flow}. 
To further enhance FM and make it effective in the DT context, e.g., for steering the virtual model toward the evolving distribution of physical parameters inferred from streaming measurements, we incorporate a confidence-adaptive weighting mechanism into the training process.   
Specifically, we replace the standard average loss in FM with a weighted loss derived through a meta-learning framework \cite{ren2018learning}. In this setting, the weights are dynamically updated at each training iteration using a small validation set as guidance. This reweighting strategy biases training toward parameter regimes that are most representative of the current system state, thereby tightly coupling generative dynamics with online calibration. By continuously adapting sample weights, the model becomes more robust to class imbalance and noisy labels, and more resilient to biased training data \cite{ren2018learning}.
This property is essential for ensuring reliable DT performance under distribution shift, sensor degradation, and unforeseen operating conditions, particularly in high-stakes spaceborne applications such as on-orbit capture, docking, and debris removal \cite{flores2014review}. In this context, a substantial body of research has focused on methods for estimating the inertial properties of servicing spacecraft and their robotic systems in support of these operations \cite{meng2019identification}. Accurate estimation of a spacecraft’s inertia parameters, attitude and position has been pursued, e.g., through Kalman Filtering \cite{lichter2003estimation}. These techniques typically rely on gyroscope measurements, which are  available only when dealing with cooperative targets. 

Despite the growing research interest in DT technology, there remains a lack of studies demonstrating how advanced generative learning frameworks can effectively address uncertainty, improve estimation accuracy, and ensure physical consistency of DTs. To address these challenges, we develop and validate a new weighted flow matching framework integrated within a DT architecture to enhance accuracy, stability, and parameter estimation under uncertainty and noisy sensing conditions. 
The new approach can differentiate between reliable and noisy samples during training, while offering a pathway to enhance estimation accuracy and robustness for inertia and state estimation.
The main contributions of this paper are: 
\begin{itemize}
\item Development of a new weighted flow matching generative framework capable of integrating confidence-adaptive  reweighting into model training; \vspace{0.2cm}

\item Development of a physics-informed filtering architecture that couples WFM with the Unscented Kalman Filter (UKF) to improve parameter estimation in DTs. We refer to this new nonlinear filtering approach as ``Boosted UKF''; \vspace{0.2cm}

\item Establish robustness, stability, and convergence of the proposed method under various excitation regimes and noise conditions; \vspace{0.2cm}

\item Demonstrate the effectiveness of the integrated framework within a DT architecture for accurate and stable inertia estimation in spacecraft systems. 
\end{itemize}

This paper is organized as follows. In Section~\ref{sec:methodology}, we present the key components of the proposed Boosted UKF and their integration within the DT architecture. In Section~\ref{sec:experiments}, we describe the application of the Boosted UKF to a rigid-body spacecraft inertia estimation problem. This includes a thorough description of data generation, classification, model training, excitation regimes, and the integration of the LRW, WFM,  and UKF components. In Section~\ref{sec:results_discussion}, we compare the proposed Boosted UKF with the UKF,  the Extended Kalman Filter (EKF),  the Ensemble Kalman Filter (EnKF), and demonstrate the superior performance of the Boosted UKF, particularly in excitation--limited regimes. Finally, in Section~\ref{sec:conclusion}, we summarize the main findings and discuss their broader implications. We also include two brief appendices providing a brief overview of the UKF, flow matching and its variants (including the proposed WFM approach).

\section{Problem formulation and new  algorithm}
\label{sec:methodology}

In this section, we present the key components of the proposed generative 
nonlinear filtering approach, i.e., the Boosted UKF, for enhanced parameter 
estimation under uncertainty and noisy sensing conditions.

\subsection{Model equations}
\label{sec:training-setup}

We model the DT as a discrete-time dynamical system 
\begin{equation}
{x}_{k+1} = f\left(x_k, \tau_k,\Delta t\right)+\eta_k, \quad x_k \in\mathbb{R}^{n}, \quad k=0,1, \ldots,
\label{U1}
\end{equation}
where $x_k$ denotes the state vector at time $t_k=k\Delta t$, $\Delta t$ is the time step size, 
$\tau_k$ is a forcing vector at time $t_k$, $f(\cdot)$ is a nonlinear state-transition function (mapping), and 
$\eta_k$ is i.i.d. zero mean Gaussian noise with covariance $Q_k$. We treat $f(\cdot)$ as a given nonlinear discrete-time state-transition mapping that advances the state from $t_k$ to $t_{k+1}$ over one $\Delta t$ step. The state vector $x_k$ includes both the system physical state and the system parameters, which are incorporated through standard state-augmentation techniques, i.e., 
\begin{equation}
x_k =
\begin{bmatrix}
\xi_k\\
\theta_k
\end{bmatrix},
\qquad
\xi_k \in \mathbb{R}^{n_s},\ \theta_k \in \mathbb{R}^{n_\theta},\ n_s+n_\theta = n,
\label{eq:aug_state_partition}
\end{equation}
where $\xi_k$ denotes the physical (dynamic) state at time $k$, and $\theta_k$ denotes the (unknown) parameters to be estimated. We are also given noisy measurements of a phase-space function $h(\cdot)$ 
\begin{equation}
z_k = h(x_k) + v_k,
\label{M1}
\end{equation}
where $h:\mathbb{R}^{n}\rightarrow\mathbb{R}^{m}$ is the observation operator, and $v_k\in\mathbb{R}^{m}$ are i.i.d. Gaussian vectors, with covariance $R_k$. Here, $h(\cdot)$ is assumed known; in general, it may depend on the full augmented state, but in our setup it represents the physical sensor model and  measurements of the physical-state components only. We are interested in estimating the system 
parameters $\theta_k$, using \eqref{U1}-\eqref{M1}. 

\begin{figure*}[!t]
  \centering
  \includegraphics[width=\textwidth]{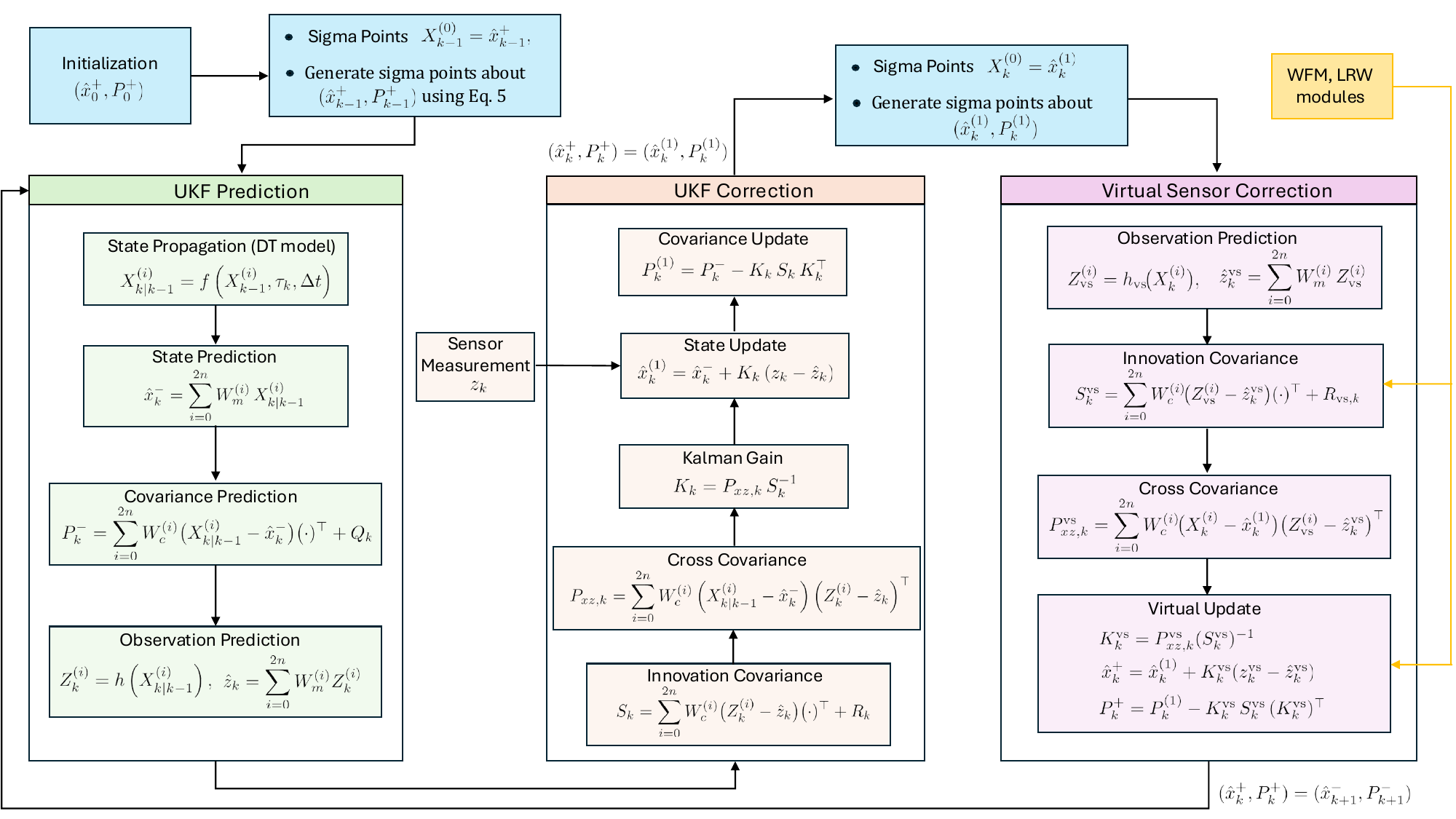}
 
  \vspace{2mm}
  \begin{minipage}{0.9\textwidth}
    \centering
    \caption{Flow chart of Boosted UKF: Parameter Identification Pipeline with Virtual-Sensor Correction.}
    \label{fig:flowchart}
  \end{minipage}
\end{figure*}

\subsection{Proposed Boosted UKF Algorithm}
\label{sec:boostedUKF}

The flow chart in Figure~\ref{fig:flowchart} summarizes the 
full parameter identification pipeline via the proposed Boosted 
UKF algorithm in four sequential stages: 
\begin{enumerate}

\item[(i)] {\em Preprocessing:}
The LRW-guided WFM generative model presented in Appendix~II is trained offline in a preprocessing step. This step consists of processing samples trajectories of \eqref{U1}, reweighting them using the LRW Algorithm~\ref{alg:LRW}, and subsequently training the WFM model via Algorithm~\ref{alg:WeightedFlow}. 
We emphasize that, while the WFM algorithm is an important 
contribution of this paper, its  presentation is deferred 
to Appendix~II to maintain the focus on nonlinear filtering.

\item[(ii)] {\em Initialization:} A design-informed 
Gaussian prior with mean $\hat{x}_0^+$ and 
covariance $P_0^+$ is 
used to initialize the filter.
\item[(iii)]  {\em UKF prediction--correction:}
The UKF algorithm and associated material are standard and summarized in Appendix~I. At each time step, the algorithm generates sigma points about the current posterior, propagates them through the discretized process model, and computes the predicted mean, covariance, and measurement.
The sensor update then uses the innovation covariance, cross-covariance, and Kalman gain to produce the corrected posterior $(\hat{x}_k^{(1)}, P_k^{(1)})$. 
See the UKF prediction and correction blocks in Figure~\ref{fig:flowchart}.
\item[(iv)] {\em Virtual-sensor correction:}
This step constitutes the main innovation over the standard UKF algorithm. Sigma points are rebuilt about $(\hat{x}_k^{(1)}, P_k^{(1)})$ and mapped through the parameter observation $h_{\mathrm{vs}}(x)$. The LRW-guided WFM statistics (see Appendix~II) are introduced as a pseudo-measurement characterized by $(\mu_{\mathrm{wfm}}, \Sigma_{\mathrm{wfm}})$. This information is incorporated through a second Kalman update, denoted as ``Virtual Sensor Correction'' in Figure~\ref{fig:flowchart}, yielding the final posterior $(\hat{x}_k^{+}, P_k^{+})$, which initializes the subsequent prediction step.
\end{enumerate}

The virtual-sensor correction block is motivated by excitation-limited regimes, in which the parameters are only weakly identifiable from measurements and standard nonlinear filters may drift or exhibit bias. The virtual sensor update leverages predictions from a Weighted Flow Matching generative model trained on a reliability-weighted dataset to infer an offline distribution over the parameters of interest. After training the WFM model offline, we draw samples from the learned flow and approximate the resulting non-Gaussian parameter distribution by a Gaussian with mean $\mu_{\mathrm{wfm}}$ and covariance $\Sigma_{\mathrm{wfm}}$. This Gaussian distribution is then used as a prior to regularize the filter toward physically consistent parameter regimes, which is beneficial in excitation--limited settings.
The WFM algorithm we propose is described in Appendix~II-\ref{sec:weighted-fm}, and is incorporated into the virtual-sensor correction block through the pseudo-measurement $(\mu_{\mathrm{wfm}}, \Sigma_{\mathrm{wfm}})$.
Specifically, we define
\begin{equation}
z_k^{\mathrm{vs}} = \mu_{\mathrm{wfm}}, \qquad 
R_{\mathrm{vs},k} = \Sigma_{\mathrm{wfm}}.
\label{eq:lrw/wfm}
\end{equation}
Sigma points are then constructed about the current posterior estimate $(\hat{x}_k^{(1)}, P_k^{(1)})$, and used to form the predicted virtual measurement and its associated covariance (see Appendix~I).
Define the virtual measurement sigma points as
\begin{equation}
Z_{\mathrm{vs}}^{(i)} = h_{\mathrm{vs}}\!\big(X_k^{(i)}\big), \qquad i=0,1,\dots,2n,
\end{equation}
where  $h_{\mathrm{vs}}$ takes the part of $X_k^{(i)}$ that defines the parameters of the model. 
Then
\begin{equation}
\hat{z}_k^{\mathrm{vs}} = \sum_{i=0}^{2n} W_m^{(i)}\, Z_{\mathrm{vs}}^{(i)}.
\end{equation}
The corresponding innovation covariance and cross covariance are
\begin{equation}
S_k^{\mathrm{vs}} = \sum_{i=0}^{2n} W_c^{(i)}\!\big(Z_{\mathrm{vs}}^{(i)}-\hat{z}_k^{\mathrm{vs}}\big)
\big(Z_{\mathrm{vs}}^{(i)}-\hat{z}_k^{\mathrm{vs}}\big)^\top + R_{\mathrm{vs},k},
\end{equation}
\begin{equation}
P_{xz,k}^{\mathrm{vs}} = \sum_{i=0}^{2L} W_c^{(i)}\!\big(X_k^{(i)}-\hat{x}_k^{(1)}\big)
\big(Z_{\mathrm{vs}}^{(i)}-\hat{z}_k^{\mathrm{vs}}\big)^\top.
\end{equation}
The virtual Kalman gain and update equations are
\begin{equation}
K_k^{\mathrm{vs}} = P_{xz,k}^{\mathrm{vs}}(S_k^{\mathrm{vs}})^{-1},
\end{equation}
\begin{equation}
\begin{cases}
\begin{aligned}
\hat{x}_k^{+} &= \hat{x}_k^{(1)} + K_k^{\mathrm{vs}}\!\left(z_k^{\mathrm{vs}}-\hat{z}_k^{\mathrm{vs}}\right),\\
P_k^{+} &= P_k^{(1)} - K_k^{\mathrm{vs}}\,S_k^{\mathrm{vs}}\,(K_k^{\mathrm{vs}})^\top.
\end{aligned}
\end{cases}
\label{eq:update}
\end{equation}
Here, $(\hat{x}_k^{(1)}, P_k^{(1)})$ are obtained from the real-sensor update in equation~\eqref{eq:update11}. If the virtual sensor update is not applied, the posterior remains equal to the real sensor posterior,
and 
$\hat{x}_k^{+},P_k^{+}=\hat{x}_k^{(1)},P_k^{(1)}$.
The resulting posterior $(\hat{x}_k^{+}, P_k^{+})$ initializes the next prediction step, ensuring a consistent Bayesian propagation of the filter
state.

\section{Application to spacecraft dynamics}
\label{sec:experiments}
We demonstrate the proposed Boosted UKF in a rigid-body 
spacecraft inertia estimation case study. The code for this 
experiment is publicly available at~\cite{BoostedUKFrepo}.
As is well known, the rotational motion of a spacecraft is governed by Euler’s 
equations \cite{meng2019identification, fernandes2024aircraftdynamics}, 
which can be expressed in the body-fixed frame $\mathcal{C}$ relative to an 
inertial frame $\mathcal{I}$ as
\begin{equation}
\label{eq:Torque}
J_{\mathcal C}
\frac{d}{dt}\!\left(
  \omega_{\mathcal C/\mathcal I}
\right)\!\Big|_{\mathcal C}
=\;
-\,
\omega_{\mathcal C/\mathcal I}
\times
\left(
  J_{\mathcal C}\,\omega_{\mathcal C/\mathcal I}
\right)
+\,
\tau_{\mathcal C}.
\end{equation}
Here  $J_{\mathcal{C}} = \mathrm{diag}(J_{x}, J_{y}, J_{z}) \in \mathbb{R}^{3\times 3}$ denotes the principal moments of inertia relative to body frame $\mathcal{C}$, 
${\omega}_{\mathcal{C}/\mathcal{I}} \in \mathbb{R}^3$ the body 
angular velocity, and ${\tau}_{\mathcal{C}} \in \mathbb{R}^3$ the applied external torque. 
%
%
The spacecraft attitude is represented by a unit quaternion
$q_{\mathcal{I}/\mathcal{C}}=[\,q_w,q_x,q_y,q_z\,]^\top\in\mathbb{R}^4$ 
with $\|q_{\mathcal{I}/\mathcal{C}}\| = 1$, governed by the equation
\begin{equation}
\label{eq:quaternion}
\dot q_{\mathcal{I}/\mathcal{C}} = \frac{1}{2}\,
q_{\mathcal{I}/\mathcal{C}} \otimes 
\begin{bmatrix} 0 \\ \omega_{\mathcal{C}/\mathcal{I}} \end{bmatrix},
\end{equation}
where $\otimes$ denotes the standard quaternion multiplication. 
The ODE system \eqref{eq:Torque} is discretized using RK4 with step size $\Delta t$ to obtain a system in the form \eqref{U1}. Here and throughout, $\|\cdot\|$ denotes the Euclidean $2$-norm for vectors. To emulate realistic sensor data, synthetic measurements are generated by corrupting the simulated baseline with Gaussian noise. Specifically,
\begin{equation}
\label{eq:11}
\omega_k = \omega_{\mathrm{hist},k} + \eta_{\omega,k},
\qquad
\eta_{\omega,k} \sim \mathcal{N}\!\big(0,\sigma_{\mathrm{gyro}}^{2} I_{3}\big),
\end{equation}
and
\begin{equation}
q_k = q_{\mathrm{hist},k} + \eta_{q,k},
\qquad
\eta_{q,k} \sim \mathcal{N}\!\big(0,\sigma_{\mathrm{quat}}^{2} I_{4}\big),
\end{equation}
with $q_k$ renormalized to unit length. The measurement vector is
\begin{equation}
\label{eq:12}
z_k = \big[q_k^\top,\ \omega_k^\top\big]^\top \in \mathbb{R}^7.
\end{equation}
Equations~\eqref{eq:11}--\eqref{eq:12} thus define the noisy measurement vector $z_k$ used in the measurement model~\eqref{M1}. The UKF is initialized using the first measurement for the state components, with covariances set by $\sigma_{\mathrm{quat}}$ and $\sigma_{\mathrm{gyro}}$, and with a broad design-informed prior for the unknown parameters. WFM samples are not used at initialization; instead, the prior statistics $(\mu_{\mathrm{wfm}},\Sigma_{\mathrm{wfm}})$ are incorporated later through the virtual-sensor update. To produce training data for LRW and WFM, we generate trajectories from~\eqref{eq:Torque} with no external torque ($\tau_\mathcal{C}\equiv 0$) through the following steps:
\begin{enumerate}
    \item {\em Sampling of inertia:} Random draws of \((J_{x},J_{y},J_{z})\) are inserted into~\eqref{eq:Torque} and integrated numerically to generate reference angular velocity trajectories \(\omega^{*}(t)\), for $t$ within a specified time interval $[0,T]$.
    \item {\em Noise injection:} Gaussian perturbations are applied to \(\omega^{*}(t)\) for $t$ in $[0,T]$, producing measured signals \(\omega_{m}(t)\) of varying fidelity, emulating sensors with different noise levels.
    \item {\em Synthetic references:} Independent inertia samples are used to compute simulated trajectories \(\omega_{s}(t)\) for reliability assessment.
    \item {\em Reliability scoring:} An error metric \(\|\omega_{s}(t)-\omega_{m}(t)\|\) quantifies the agreement between simulated and measured signals. Low-error cases are deemed reliable and form a clean meta set, while the broader training set includes both reliable and noisy samples. 
    \item {\em Learning to reweight (Algorithm \ref{alg:LRW}):} LRW assigns adaptive weights \(w_i\) to training samples, guided by the meta set, so that physically consistent data are emphasized while outliers are down-weighted.
    \item {\em Weighted flow matching (Algorithm \ref{alg:WeightedFlow}):} The reweighted dataset is passed to WFM, which trains a generative vector field that transports a Gaussian prior to the distribution of angular velocities aligned with reliable data.
\end{enumerate}
Through this sequence of steps, LRW identifies trustworthy samples, enabling WFM to learn a generative prior that more faithfully captures the spacecraft’s inertial dynamics. The resulting reliability-aware priors for $\omega(t)$ reflect the underlying dynamics without reintroducing the full measurement model or relying on torque-driven excitation. These priors are subsequently fused with onboard measurements through the UKF to refine both state and inertia estimates.

\subsection{Data generation}
To test the proposed Boosted UKF, we consider the following 
inertia matrix (see \cite{basu2025hybrid}).
\begin{equation}
\label{eq:Jtrue}
J_{\mathcal C}^{\star}=\begin{bmatrix}
100 & 0 & 0\\
0 & 80 & 0\\
0 & 0 & 70
\end{bmatrix}\ \mathrm{[kg\cdot m^2]}.
\end{equation}
This matrix satisfies the  well-known triangle inequalities on its principal moments, and is never made available to the algorithm at any stage. We will refer to $J_{\mathcal C}^{\star}$ as the ``true'' inertia matrix. Our goal is to estimate $J_{\mathcal C}^\star$. We initialize the angular velocity as
\[
\omega(0) = [0.1,\,0.1,\,0.1]\ \mathrm{rad\,s^{-1}}.
\]
Using $J_{\mathcal C}^\star$ together with this initial condition, we generate a reference trajectory $\omega^\ast(t)$ by numerically integrating \eqref{eq:Torque} over the time interval $t \in [0,30]\ \mathrm{s}$.
We then collect $M = 600$ uniformly spaced temporal snapshots of $\omega^\ast(t)$ over $t \in [0,30]\ \mathrm{s}$, at times
\begin{equation}
t_k = k \Delta t, \quad k=1,2, \ldots, M, \quad \Delta t = 0.05\ \mathrm{s}.
\label{tgrid}
\end{equation}

\begin{figure}[!t]
  \centering
  \includegraphics[width=.90\columnwidth]{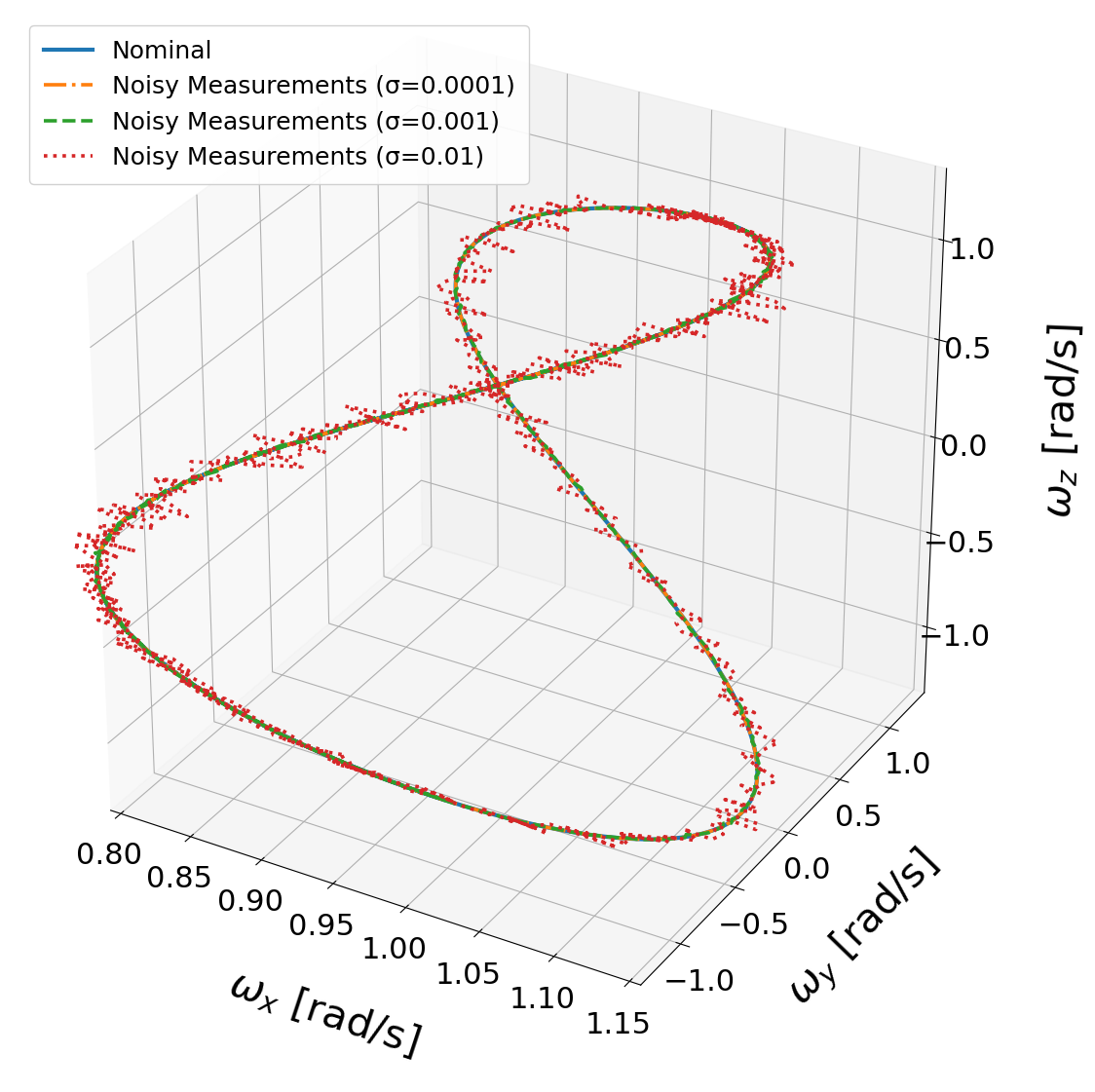} 
  \caption{Angular velocity nominal trajectory and simulated noisy measurements.}
  \label{fig:traj_samples}
\end{figure}

To emulate sensor uncertainty, zero-mean Gaussian noise with standard deviations $\sigma\in\{10^{-4},\,10^{-3},\,10^{-2}\}\ \mathrm{rad\,s^{-1}}$ is added to produce noisy observations $\omega_{m}(t)$. 
To represent model mismatch, a set of $N=2000$ diagonal inertia matrices is drawn at random by perturbing each principal moment with a $10\%$ Gaussian relative deviation
\begin{equation}
\begin{aligned}
J_{x} &\sim \mathcal{N}(100,\,10^{2}),\\
J_{y} &\sim \mathcal{N}(80,\,8^{2}),\\
J_{z} &\sim \mathcal{N}(70,\,7^{2}).
\end{aligned}
\end{equation}
For each sample $J_{s}$ of the matrix of inertia, equation~\eqref{eq:Torque} is integrated numerically to obtain a surrogate trajectory $\omega_{s}(t)$, $t\in [0,30]\,\mathrm{s}$. The degree of agreement with the noisy observation is quantified by the discrete mean-squared error
\begin{equation}
e_s =\frac{1}{M}\sum_{k=1}^{M}\left\| \omega_{s}(t_k)-\omega_{m}(t_k)\right\|^{2}, \quad s=1,2, \ldots,N, 
\label{errorLRW}
\end{equation}
where $t_k$ is the temporal grid defined in \eqref{tgrid}. 
The error \(e_s\) is mapped to per-sample weights the LRW Algorithm \ref{alg:LRW} discussed in Appendix II-\ref{sec:non-linear-weights}. These reliability-aware weights rebalance the dataset, ensuring that WFM focuses on trajectories consistent with the dynamics. In Figure~\ref{fig:traj_samples} we show the nominal trajectory ${\omega}^{\ast}(t)$, $t\in[0,30]\,\mathrm{s}$, a surrogate trajectory ${\omega}_{s}(t)$ obtained with a perturbed inertia, and three 
measured trajectories ${\omega}_{m}(t)$. It is seen that as the noise level increases, the measured 
trajectories deviate more strongly from the nominal (``true'') trajectory. 

\begin{table*}[!t]
  \centering
  \small
  \setlength{\tabcolsep}{5pt}
  \renewcommand{\arraystretch}{1.15}

 \begin{tabular}{|m{100pt}|m{150pt}|m{175pt}|}
    \hline
    \textbf{Parameter} & \textbf{Value} & \textbf{Description / Units} \\
    \hline
    Prior inertia mean $\mu$ &
    $[140,\,20,\,36]$ &
    Vector representing the mean of the principal moments of inertia in $\mathrm{kg\cdot m^2}$ \\
        \hline
    Prior inertia covariance $\Sigma$ &
    $\operatorname{diag}\,(1700,\,20,\,120)$ &
    Prior inertia covariance at initial time $\mathrm{kg^2\cdot m^4}$ \\
         \hline
    $ \hat{x}_0^+$&
    $\bigl[\,q_0 ,\, \omega_0,\,\mu \ \bigr]$ &
    $q_0=[1,\,0,\,0,\,0]$, \quad    $\omega_0=[0.1,\,0.1,\,0.1]\ \mathrm{rad/s}$. \\
    \hline
    $P^+_0$ &
    $\operatorname{diag}\,\!\bigl(10^{-3}I_4,\,10^{-2}I_3,\,\Sigma\bigr)$ &
    State covariance; quaternion block, angular-rate block $(\mathrm{rad/s})^{2}$, and inertia block. $I_n$ denotes the $n \times n$ identity matrix \\
    \hline
    $Q_0$ &
    $10^{-7} \operatorname{diag}\,\!\bigl(I_4,\,I_3,\,I_3\bigr)$ &
    Process-noise covariance; quaternion block, angular-rate block $(\mathrm{rad/s})^{2}$, and inertia block \\
    \hline
    $R_0$ &
    $2.5\times 10^{-5} \operatorname{diag}\,\!\bigl(I_4,\,I_3\bigr)$ &
    Measurement-noise covariance; quaternion sensor $\sigma_{\text{quat}}=0.005$, gyro $\sigma_{\text{gyro}}=0.005\ \mathrm{rad/s}$ \\
    \hline
  \end{tabular}
   \caption{Initialization parameters for the Boosted UKF algorithm described in Section~\ref{sec:boostedUKF}.} 
  \label{tab:enkf_settings}
\end{table*}

\vspace{0.5cm}
\begin{figure*}[!t]
\centerline{\hspace{-0.4cm}PDF of re-weighted training samples\hspace{1.5cm} Weighted flow matching prediction}
\vspace{0.3cm}
  \centering

  \subfloat[Sensor-noise level $\sigma = 10^{-2}\,\mathrm{rad\,s^{-1}}$.]{%
    \includegraphics[width=0.75\textwidth]{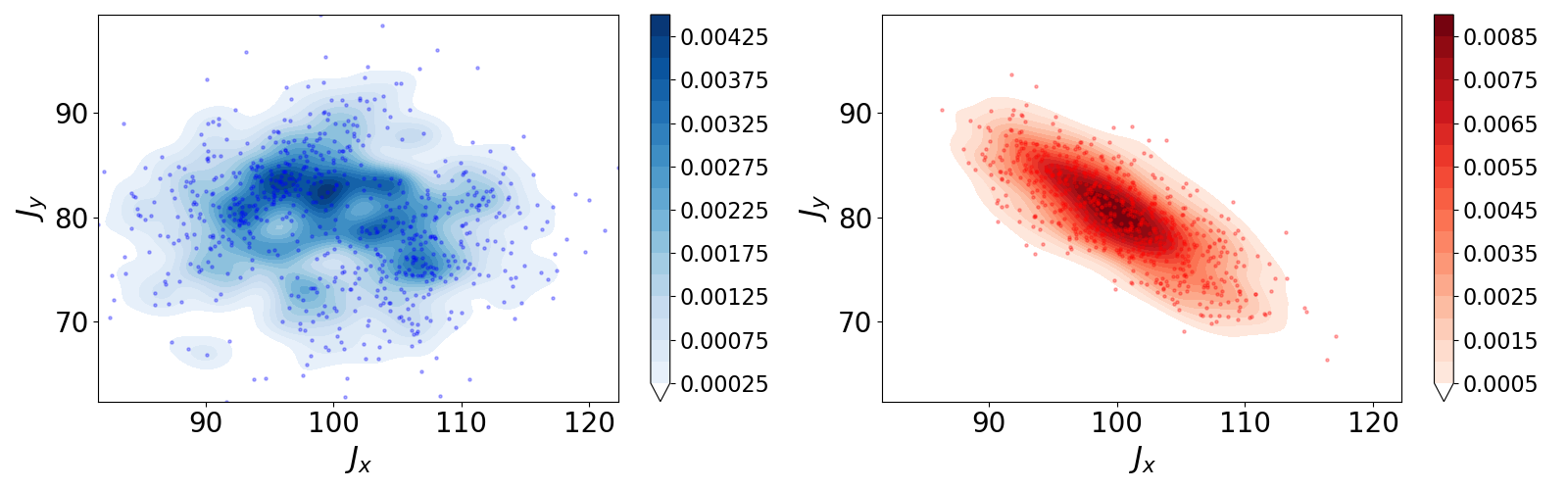}%
    \label{fig:wfm_noise_1e-2}
  }\\[5pt]

  \subfloat[Sensor-noise level $\sigma = 10^{-3}\,\mathrm{rad\,s^{-1}}$.]{%
    \includegraphics[width=0.75\textwidth]{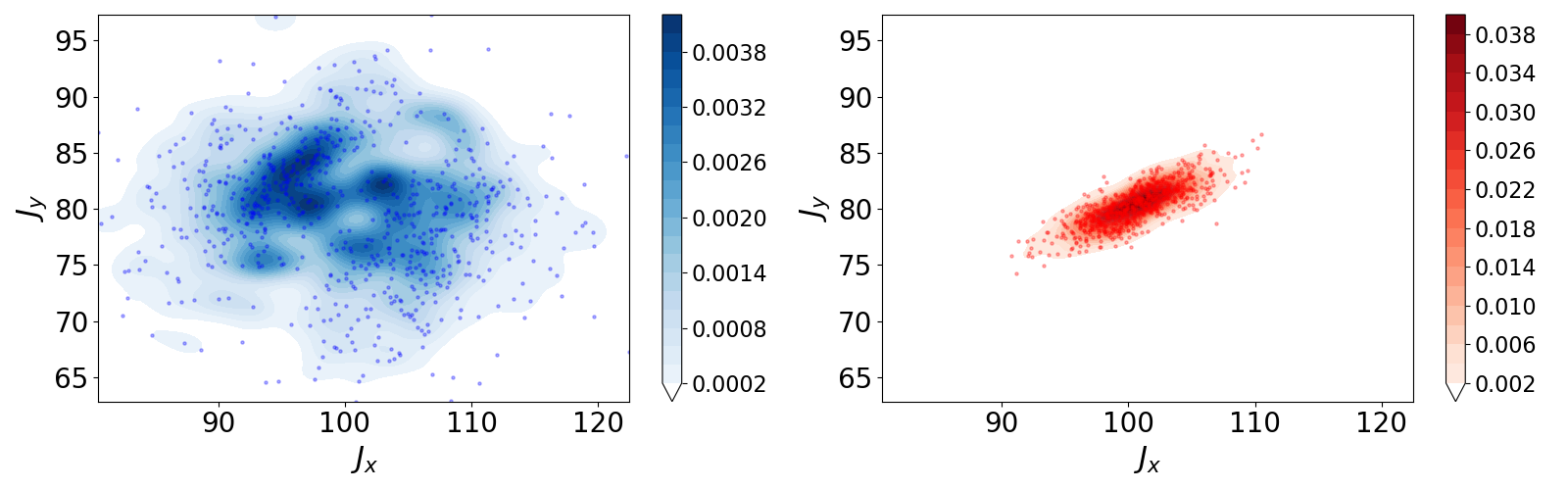}%
    \label{fig:wfm_noise_1e-3}
  }\\[5pt]

  \subfloat[Sensor-noise level $\sigma = 10^{-4}\,\mathrm{rad\,s^{-1}}$.]{%
    \includegraphics[width=0.75\textwidth]{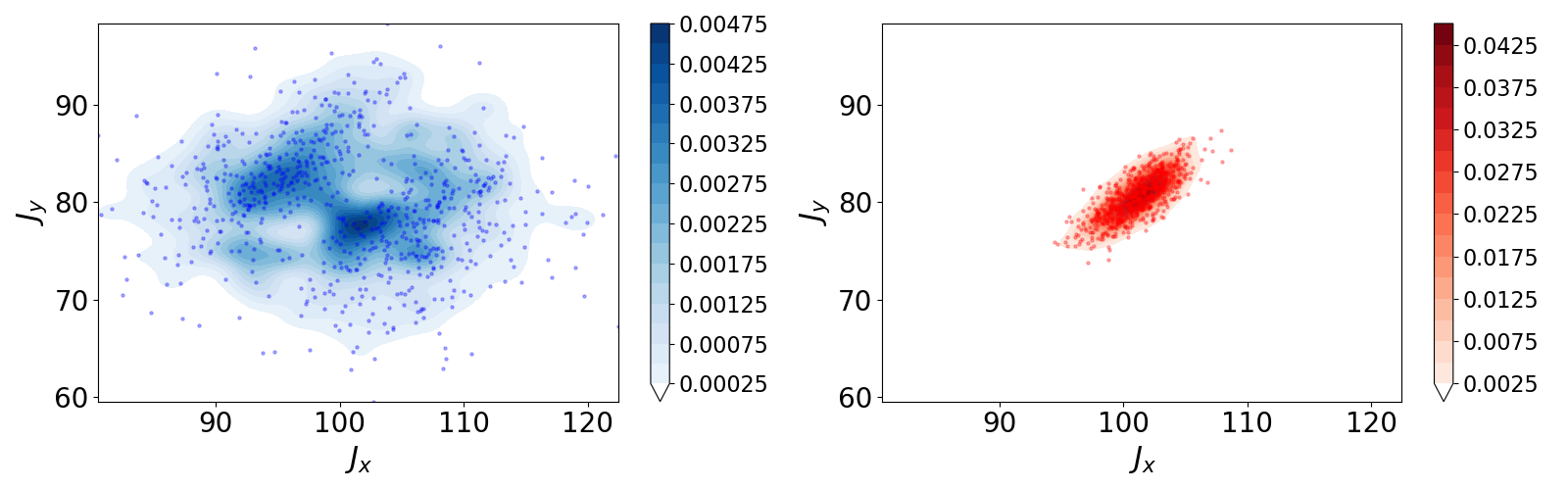}%
    \label{fig:wfm_noise_1e-4}
  }

  \vspace{2mm}
  \begin{minipage}{\textwidth}
    \centering
    \caption{PDF of re-weighted training samples (left panel), and PDF of WFM-generated samples (right panel). We show results for different sensor-noise levels $\sigma$. It is seen that as $\sigma$ decreases WFM learns a distribution that concentrates progressively closer to the unknown nominal inertia matrix \eqref{eq:Jtrue}.}
    \label{fig:wfm_kde_stack}
  \end{minipage}
\end{figure*}

\subsection{Learning to re-weight (LRW)} 
\label{sec:mainLWR}
The LRW module is implemented with a shallow Multi-Layer Perceptron (MLP). 
The input consists of a single scalar feature, namely the $z$-score of the angular velocity error, defined as $ z_s = (e_s - \mu_e)/\sigma_e$, where $\mu_e$ and $\sigma_e$ are, respectively, the mean and  standard deviation of the vector $e$ with components \eqref{errorLRW}. 
This feature is processed by two fully connected hidden layers 
with 64 and 32 neurons, respectively, each followed by ReLU activation, and a final output 
layer producing a scalar logit that represents the reliability weight of each sample. 
Ground-truth labels indicate whether a trajectory belongs to the low-error (class~0) or 
high-error (class~1) group. Training is performed for 200 epochs using 
stochastic gradient descent. The baseline model 
uses a fixed learning rate of \(10^{-4}\), while the meta-learning model uses 
\(3 \times 10^{-4}\), both with a batch size of 32. During meta-learning, a bilevel optimization 
framework is used: in the inner loop, perturbation variables \(\epsilon\) are introduced to 
assign sample-specific weights based on their influence on the meta-validation loss. These weights 
are constrained to nonnegative values and normalized via $\ell_1$-scaling within each batch to form a valid distribution. After each epoch, training and test accuracy are recorded to 
compare the baseline and meta-learning configurations. At convergence, the learned 
weights are averaged across 200 epochs, and both raw and normalized 
weight values are stored alongside their corresponding surrogate inertia parameters. 

\subsection{Weighted flow matching (WFM)}
The reweighted training data set is used to train WFM (see Appendix II-\ref{sec:weighted-fm}). The flow matching base conditional vector field $v_t(x)$ is parameterized by a
five-layer MLP with hidden widths $[256,256,256,256,256]$ and sinusoidal time 
encoding. Training runs for $10{,}000$ epochs with Adam (learning rate $10^{-5}$), minimizing a weighted FM loss (see Appendix~II for details); each sample’s contribution is scaled by its normalized weight $w_i$, which suppresses high-noise examples and emphasizes reliable ones. 
%
%
%
In Figure~\ref{fig:wfm_kde_stack}(a)-(c), we illustrate how WFM, guided by LRW, improves as the sensor
noise standard deviation $\sigma$ is reduced from $10^{-2}$ to $10^{-4}\,\mathrm{rad\,s^{-1}}$.
In each panel, the left plot shows the LRW-reweighted empirical density in the $(J_{x},J_{y})$ plane,
where darker shading indicates regions assigned higher sample weights with higher estimated reliability.
The right plot shows the corresponding kernel density estimate (KDE) of the samples generated by WFM.
As $\sigma$ decreases, the WFM-generated distribution more closely matches $J_{\mathcal{C}}^\star$, indicating improved recovery of the target parameter distribution. Lower noise yields better measurements, allowing the LRW weights to highlight trustworthy samples; WFM therefore learns a more accurate flow. This drop in bias and spread means the model can deliver tighter, more reliable inertia estimates crucial for subsequent tasks such as parameter estimation and control.

\subsection{Excitation regimes}
\label{ExcitationRegimes}
We test the Boosted UKF described in Section~\ref{sec:boostedUKF} over a
$400$\,s horizon with integration step $\Delta t=0.01$\,s.
The chief spacecraft inertia matrix was set to $J_{\mathcal C}=J_{\mathcal C}^{\star}$ from
Eq.~\eqref{eq:Jtrue}, and time histories of angular velocity
$\omega_{\text{hist}}$, attitude quaternion $q_{\text{hist}}$, and torque input
$\tau_{\text{hist}}$ were generated. Three different excitation 
regimes were applied:
\begin{enumerate}
\item [(i)] A continuous multi-frequency torque
\begin{equation}
\tau(t)=
\begin{bmatrix}
  1.0\,\sin(0.1t)+2.5\,\cos(0.3t)\\
  \quad +\,1.0\,\sin(0.7t)+1.0\,\sin(1.5t)\\[4pt]
  2.6\,\cos(0.15t)+3.0\,\sin(0.4t)\\
  \quad +\,2.4\,\cos(0.8t)+1.8\,\cos(1.8t)\\[4pt]
  3.4\,\sin(0.12t)+2.1\,\cos(0.5t)\\
  \quad +\,1.0\,\sin(0.9t)+1.5\,\sin(2.0t)
\end{bmatrix}\!\,
\label{eq:torque}
\end{equation}\vspace{0.2cm}
where $t\in[0,400]$ s. 
\item[(ii)] “Windowed” excitations with 1\,s pulses at 200–201\,s, 250–251\,s, and 300–301\,s using the same profile; \vspace{0.2cm}

\item[(iii)] “Persistent” configuration with fourteen identical 1\,s bursts every 25\,s from 50\,s through 375\,s.
\end{enumerate}

The Boosted UKF was initialized using the parameters in Table~\ref{tab:enkf_settings}.
The state $x_0$ stacks attitude, angular velocity, and log unit–shape inertia parameters;
the quaternion was initialized to identity.
The initial covariance $P_0$ imposes tight bounds on attitude and rates, and a broad design
prior covariance for the inertia block mapped to the log unit--shape coordinates.
The process-noise covariance $Q_k$ governs uncertainty propagation, while the measurement-noise covariance $R_k$ characterizes sensor accuracies ($\sigma_{\text{gyro}} = 0.005$ rad/s, $\sigma_{\text{quat}} = 0.005$). The Van der Merwe scaled sigma points are constructed using the parameters $(\alpha = 0.001,\ \beta = 2.0,\ \kappa = 0.0)$.
After each update, quaternions were renormalized, 
log unit--shape coordinates re-centered, and covariances 
symmetrized with a small jitter to preserve positive definiteness. 

\section{Numerical Results and Discussion}
\label{sec:results_discussion}

\begin{figure*}[!t]

\centerline{\hspace{0.13\textwidth}\textbf{Full}\hspace{0.24\textwidth}\textbf{Windowed}\hspace{0.22\textwidth}\textbf{Persistent}}

\vspace{6pt}
\begin{minipage}{0.08\textwidth}
\centering
\rotatebox{90}{\hspace{0.3cm}\textbf{EKF}}
\end{minipage}
\begin{minipage}{0.30\textwidth}
\includegraphics[width=\linewidth]{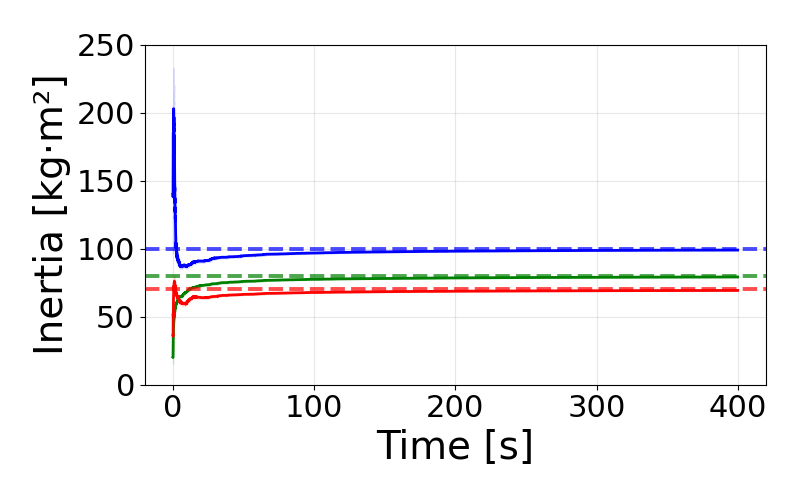}
\end{minipage}
\begin{minipage}{0.30\textwidth}
\includegraphics[width=\linewidth]{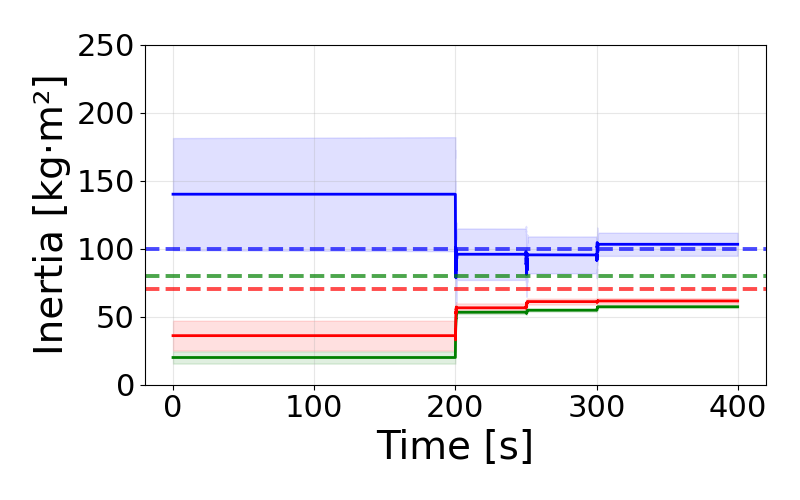}
\end{minipage}
\begin{minipage}{0.30\textwidth}
\includegraphics[width=\linewidth]{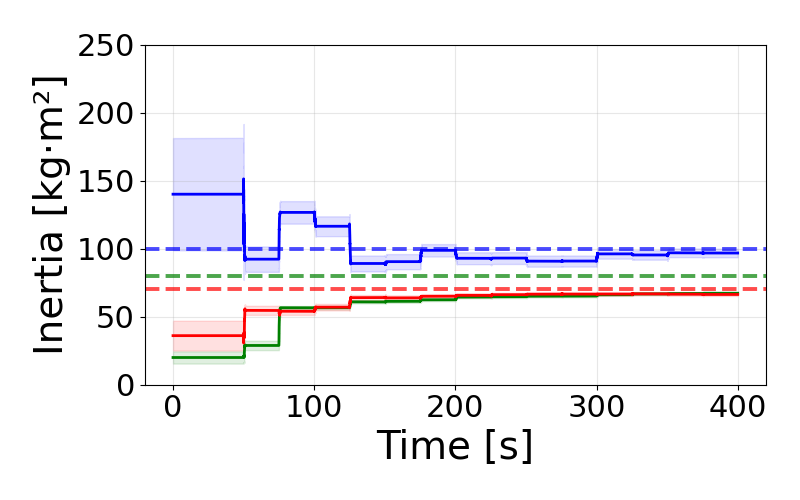}
\end{minipage}

\vspace{6pt}

\begin{minipage}{0.08\textwidth}
\centering
\rotatebox{90}{\hspace{0.3cm}\textbf{UKF}}
\end{minipage}
\begin{minipage}{0.30\textwidth}
\includegraphics[width=\linewidth]{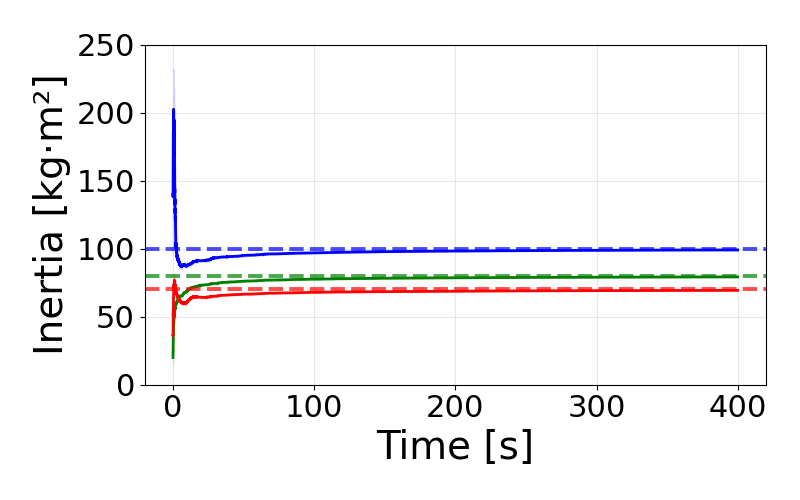}
\end{minipage}
\begin{minipage}{0.30\textwidth}
\includegraphics[width=\linewidth]{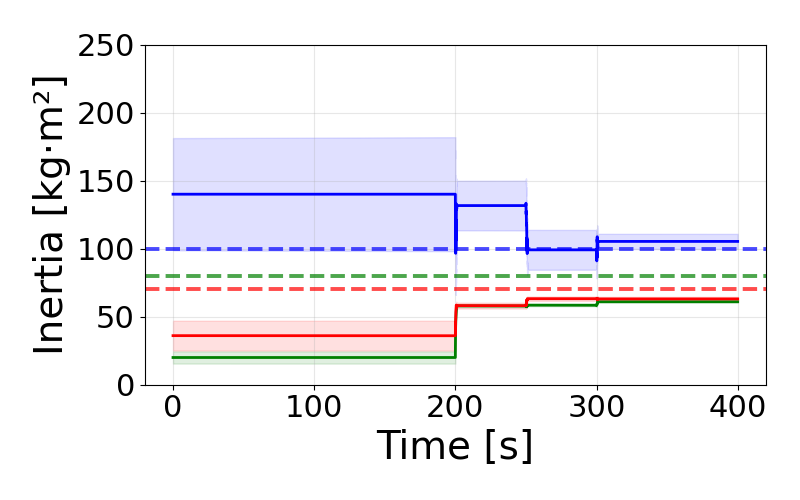}
\end{minipage}
\begin{minipage}{0.30\textwidth}
\includegraphics[width=\linewidth]{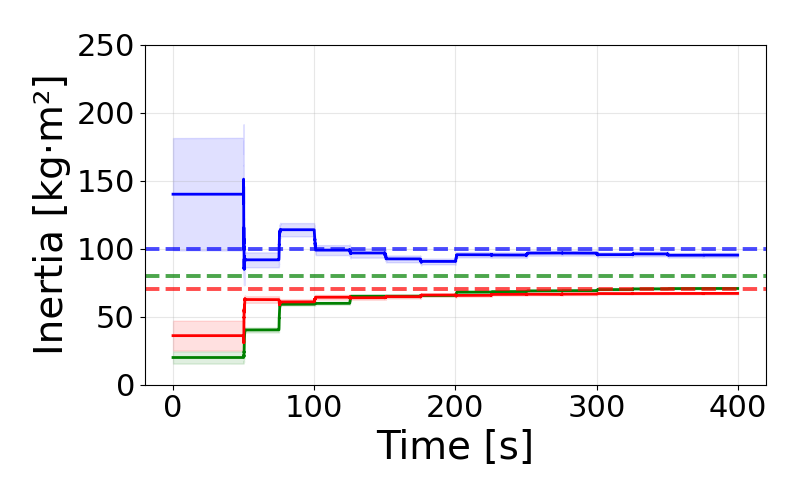}
\end{minipage}

\vspace{6pt}

\begin{minipage}{0.08\textwidth}
\centering
\rotatebox{90}{\hspace{0.4cm}\textbf{EnKF}}
\end{minipage}
\begin{minipage}{0.30\textwidth}
\includegraphics[width=\linewidth]{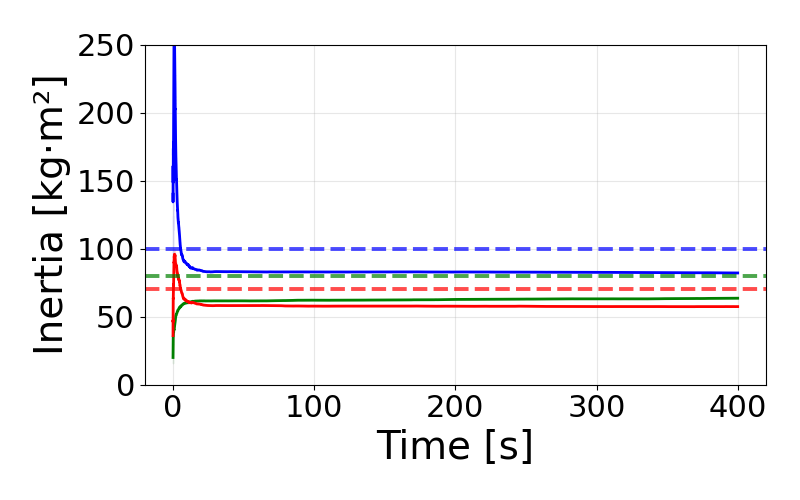}
\end{minipage}
\begin{minipage}{0.30\textwidth}
\includegraphics[width=\linewidth, height=3.1cm]{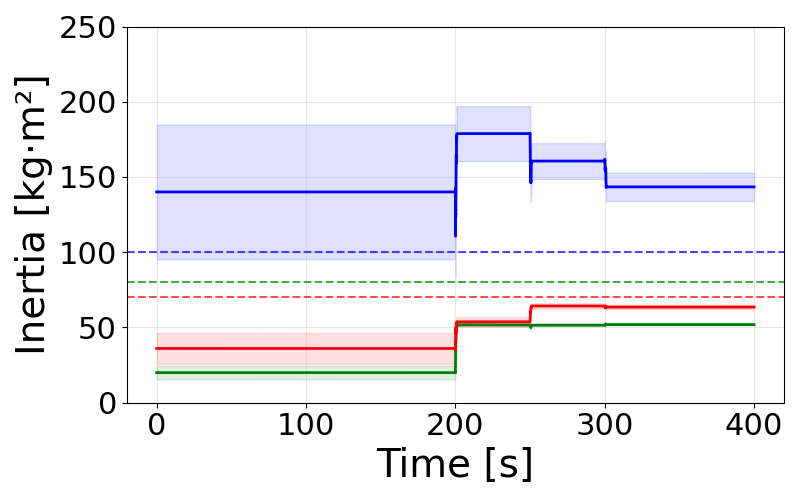}
\end{minipage}
\begin{minipage}{0.30\textwidth}
\includegraphics[width=\linewidth]{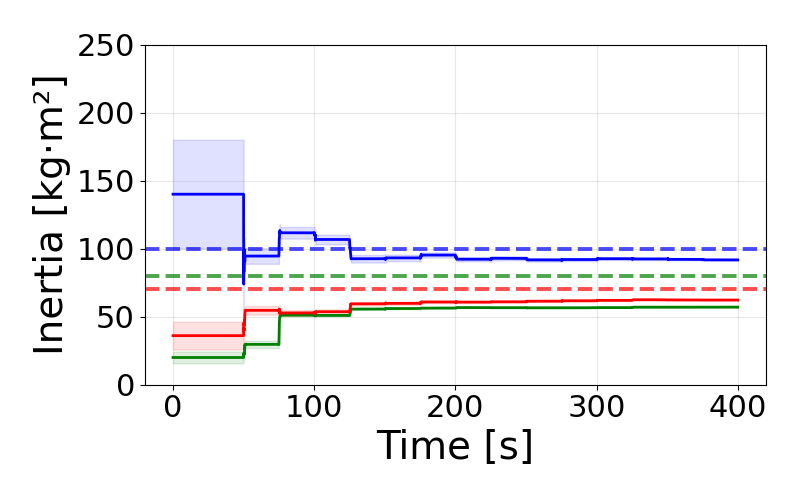}
\end{minipage}

\vspace{6pt}

\begin{minipage}{0.08\textwidth}
\centering
\rotatebox{90}{\hspace{0.5cm}\textbf{Boosted UKF}}
\end{minipage}
\begin{minipage}{0.30\textwidth}
\includegraphics[width=\linewidth]{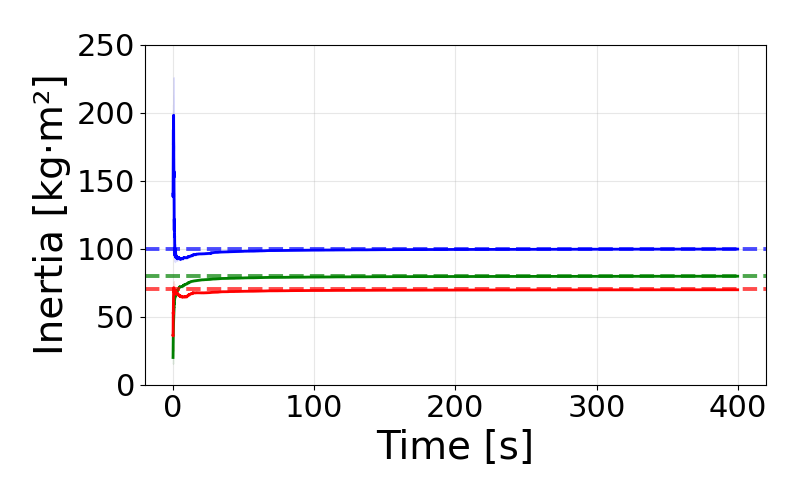}
\end{minipage}
\begin{minipage}{0.30\textwidth}
\includegraphics[width=\linewidth, height=3.0cm]{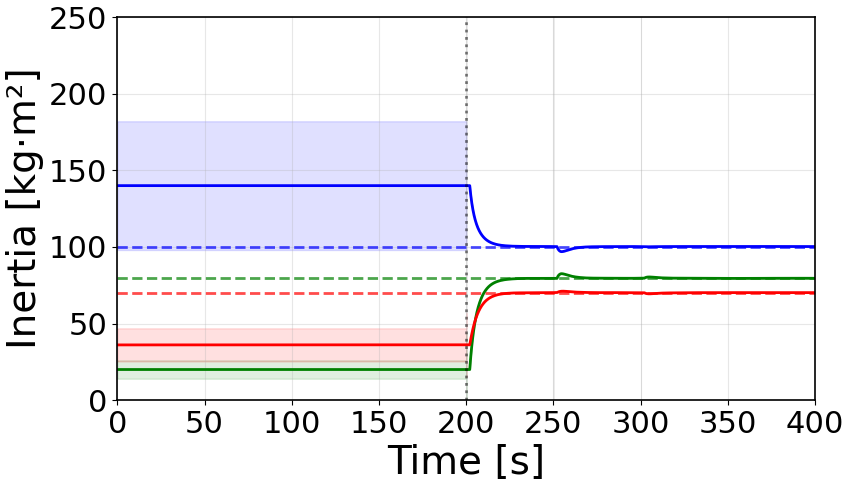}
\end{minipage}
\begin{minipage}{0.30\textwidth}
\includegraphics[width=\linewidth]{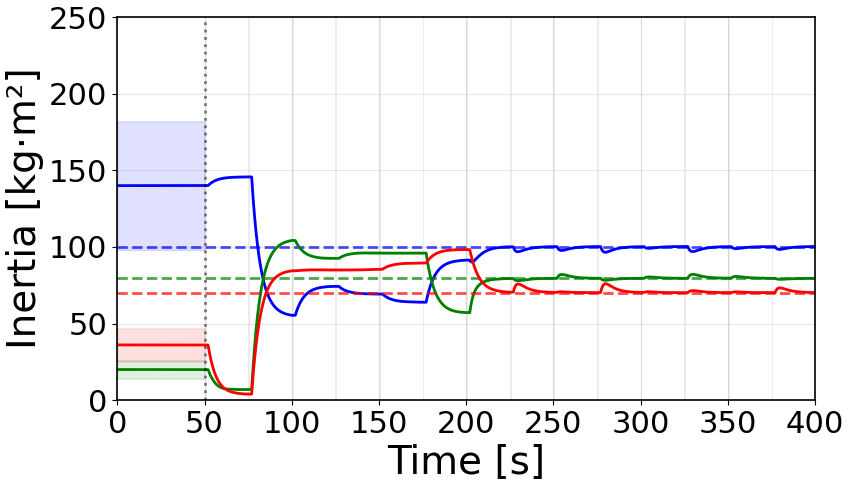}
\end{minipage}
\centerline{\hspace{2cm}
\includegraphics[width=0.7\textwidth]{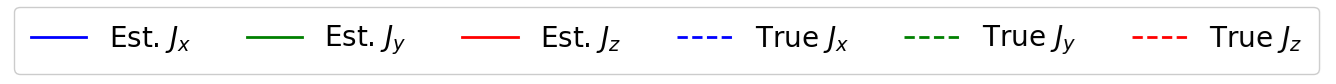}}
\caption{
Principal moments of inertia identification results under full, windowed, and persistent excitation for different nonlinear filters. Solid lines denote the mean estimates, shaded regions indicate uncertainty bands corresponding to plus or minus one standard deviation, and dashed lines represent the true values of the moments of inertia.
The results show that the proposed Boosted UKF outperforms the EKF, the UKF , and the EnKF.}
\label{fig:all_results_onepage_verticalcolumns}
\end{figure*}

In Figure~\ref{fig:all_results_onepage_verticalcolumns}, we compare the time histories of the estimated principal moments of inertia obtained using the EKF, the UKF, the EnKF, and the Boosted UKF under full, windowed, and persistent excitation. These three excitation regimes are discussed in detail in Section~\ref{ExcitationRegimes}. The results show that the proposed Boosted UKF outperforms EKF, UKF, EnKF in all excitation regimes. We now examine each case in detail.

\subsection{Full Excitation}
For full excitation (equation \eqref{eq:torque}), all four estimators converge to the nominal inertia values, albeit 
with distinct bias and uncertainty profiles. The EKF converges rapidly, yet retains a small asymptotic bias due to first-order linearization. Its final absolute errors are
\([0.18,\,0.14,\,0.13]\ \text{kg}\cdot\text{m}^2\) about \(0.18\!-\!0.19\%\) per axis, and the \(\sigma\) uncertainty bands contract to \([0.24,\,0.19,\,0.17]\ \text{kg}\cdot\text{m}^2\).
Similarly, the UKF settles at an error of \(0.05\%\) on each axis, with final spreads of \(\sigma\)
given by \([0.24,\,0.20,\,0.17]\ \text{kg}\cdot\text{m}^2\). Starting from the same coarse prior, the EnKF converges as quickly as EKF and UKF but yields an asymptotic state with larger absolute errors of approximately
\([1.42,\,1.35,\,1.02]\ \text{kg}\cdot\text{m}^2\). Finally, we see that the Boosted UKF, which fuses a physics-informed WFM prior through a scheduled virtual sensor update, yields fastest convergence and most accurate asymptotic results, with error down to $0.025\%$ of nominal values. 


\subsection{Windowed Excitation}
Next, we study performance of EKF, UKF, EnKF and Boosted UKF under windowed excitation for three short \(1\ \text{s}\) torque pulses at \(t=200,250,300\;\text{s}\). Because the impulses carry far less information than the continuous full case, the quality of each filter’s initial scale dominates the outcome. It is observed that the EKF settles quickly after each pulse. However, because the gain is tied to a first-order linearization about a mis-scaled prior, the filter locks onto an incorrect state. The final relative errors are \(12.0\%\), \(12.3\%\), and \(10.4\%\) for \(J_{x},J_{y},J_{z}\), respectively. The \(\sigma\) bands contract after each impulse but remain centered on a biased mean. 
Similarly, the UKF drifts further from the true values and stabilizes at errors of 
\(17.1\%\), \(17.1\%\), and \(15.6\%\) in \(J_{x}, J_{y}, J_{z}\), respectively. 
Although the uncertainty envelopes narrow after each pulse, they remain 
centered around a biased trajectory, indicating that higher-order propagation 
alone cannot compensate for poor scaling under low-information conditions.
The EnKF benefits from the ensemble’s stochastic spread. 
Even without explicit scale information this method reduces bias modestly, 
finishing at \(16.0\%\), \(15.3\%\), and \(12.7\%\) errors. Sampling diversity helps the filter explore alternative shapes, but with finite ensemble size and sparse data the absolute scale remains weakly observable and residual errors persist. On the other hand, we see that the proposed Boosted UKF with physics-informed WFM prior can estimate the moments of inertia to near truth with final errors of
\(0.02\%\), \(0.08\%\), and \(0.15\%\) on \(J_{x},J_{y},J_{z}\).
This case also shows the tightest \(\sigma\) envelopes, reflecting both improved identifiability 
and a well-centered prior. With intermittent actuation, the information content is concentrated 
in a few brief windows; between pulses the dynamics carry little to no new identifiability of the inertia scale. 
On the other hand, EKF and UKF inherit their steady-state bias from the prior and cannot recover once the final impulse has passed. Similarly, EnKF’s stochastic representation can reduce but not eliminate this bias under scarce data. However, injecting WFM information preconditions the problem by concentrating probability mass on physically consistent scales, so the limited excitation is needed to learn the remaining shape. Consequently, the Boosted UKF attains the smallest bias and uncertainty, providing a practical route to on–orbit inertia identification.

\subsection{Persistent Excitation}
Finally, we compare EKF, UKF, EnKF, and Boosted UKF 
under a rich, persistent excitation schedule. 
These repeated bursts provide substantially more information than the single–window case, so scale is well observed and the role of the prior is reduced. Uncertainty visibly contracts as information accumulates. It is seen that EKF tracks the step-like information gains and converges 
rapidly, but first-order linearization leaves a small axis-dependent bias. Final absolute errors are
\(0.17\%\), \(0.18\%\), and \(0.86\%\) with terminal \(\sigma\) bands
\([1.16,\,0.73,\,0.77]\;\text{kg}\cdot\text{m}^2\).
The larger residual on \(J_{zz}\) is consistent with the EKF’s sensitivity to local Jacobian accuracy during the highest gain bursts. Propagating sigma points through each excitation (UKF filter) removes the dominant linearization bias and yields the most accurate means in this rich data regime. 
Final absolute errors are
\(0.16\%\), \(0.20\%\), and \(0.21\%\) on \(J_{x},J_{y},J_{z}\).
The terminal \(\sigma\) spreads contract to
\([0.74,\,0.59,\,0.55]\;\text{kg}\cdot\text{m}^2\)
matching the tight envelopes seen after the last two excitation windows. 
EnKF also benefits from the abundant excitation but retains small sampling 
driven offsets relative to the UKF. 
Its final absolute errors are \(0.02\%\), \(0.43\%\), and \(0.4\%\),
while the ensemble’s log–shape \(\sigma\) (square–root diagonal) is
\([0.63,\,0.26,\,0.33]\). The Boosted UKF clearly  
outperforms all other methods. 
The small peaks observed in the time traces of the Boosted UKF 
under windowed excitation can be attributed to the fact that WFM 
and LRW are trained on data without torque profiles. This may lead 
to transient responses that temporarily drift away from nominal 
values.

\begin{table*}[!t]
  \centering
  \small
  \setlength{\tabcolsep}{3pt}
  \renewcommand{\arraystretch}{1.15}

  \begin{tabular}{|p{70pt}|p{95pt}|p{95pt}|p{95pt}|}
    \hline
    \textbf{Method} & \textbf{Full} & \textbf{Windowed} & \textbf{Persistent} \\
    \hline
    EKF &
    \begin{tabular}[c]{@{}l@{}}
    $J_x: 0.8633 \pm 0.7256$ \\
    $J_y: 0.8570 \pm 0.7200$ \\
    $J_z: 0.8548 \pm 0.7186$
    \end{tabular}
    &
    \begin{tabular}[c]{@{}l@{}}
    $J_x: 3.9205 \pm 3.0115$ \\
    $J_y: 22.6361 \pm 9.5308$ \\
    $J_z: 9.0447 \pm 6.4767$
    \end{tabular}
    &
    \begin{tabular}[c]{@{}l@{}}
    $J_x: 3.8957 \pm 1.4449$ \\
    $J_y: 10.4958 \pm 5.8081$ \\
    $J_z: 3.7799 \pm 1.8356$
    \end{tabular}
    \\
    \hline
    UKF &
    \begin{tabular}[c]{@{}l@{}}
    $J_x: 0.9063 \pm 0.4755$ \\
    $J_y: 0.9055 \pm 0.4746$ \\
    $J_z: 0.8996 \pm 0.4720$
    \end{tabular}
    &
    \begin{tabular}[c]{@{}l@{}}
    $J_x: 4.0534 \pm 2.9449$ \\
    $J_y: 20.6939 \pm 4.9444$ \\
    $J_z: 10.1735 \pm 6.3144$
    \end{tabular}
    &
    \begin{tabular}[c]{@{}l@{}}
    $J_x: 6.1471 \pm 1.3918$ \\
    $J_y: 15.8645 \pm 4.4491$ \\
    $J_z: 4.8835 \pm 2.1316$
    \end{tabular}
    \\
    \hline
    EnKF &
    \begin{tabular}[c]{@{}l@{}}
    $J_x: 16.9581 \pm 6.0611$ \\
    $J_y: 23.0409 \pm 6.5174$ \\
    $J_z: 17.5109 \pm 6.1284$
    \end{tabular}
    &
    \begin{tabular}[c]{@{}l@{}}
    $J_x: 24.2959 \pm 23.1755$ \\
    $J_y: 34.0652 \pm 9.2196$ \\
    $J_z: 13.4719 \pm 6.3626$
    \end{tabular}
    &
    \begin{tabular}[c]{@{}l@{}}
    $J_x: 8.6455 \pm 6.5385$ \\
    $J_y: 29.3366 \pm 8.2515$ \\
    $J_z: 10.3999 \pm 5.2649$
    \end{tabular}
    \\
    \hline
    Boosted UKF &
    \begin{tabular}[c]{@{}l@{}}
    $J_x: 0.6251 \pm 0.0456$ \\
    $J_y: 0.6570 \pm 0.0400$ \\
    $J_z: 0.6548 \pm 0.0486$
    \end{tabular}
    &
    \begin{tabular}[c]{@{}l@{}}
    $J_x: 0.2525 \pm 0.0402$ \\
    $J_y: 0.1984 \pm 0.0804$ \\
    $J_z: 0.2575 \pm 0.0638$
    \end{tabular}
    &
    \begin{tabular}[c]{@{}l@{}}
    $J_x: 0.2240 \pm 0.0469$ \\
    $J_y: 0.1419 \pm 0.0627$ \\
    $J_z: 0.2730 \pm 0.0756$
    \end{tabular}
    \\
    \hline
  \end{tabular}
\caption{Inertia estimation errors (20) for EKF, UKF, EnKF, and Boosted UKF under full, windowed, and persistent excitation with random initial conditions. Statistics are computed over $N_s=50$ initial condition realizations and correspond to errors at $t=400$ s . Entries report mean $\pm$ standard deviation of relative errors for  $J_x$, $J_y$, and $J_z$, expressed as percentages of nominal (``true'') values.}
\label{tab:all_filters_mc_results}
\end{table*}

\subsection{Estimation error for random initial conditions}
In previous sections, we examined the performance of EKF, UKF, EnKF, and Boosted UKF 
under different forcing scenarios for a fixed initial condition. We now assess robustness 
with respect to variability in the initial condition. To this end, we conduct a Monte Carlo 
study for all four filters by introducing random perturbations of the initial moment of 
inertia around the same nominal values used throughout the paper. Specifically, the diagonal 
entries of the inertia matrix are sampled from Gaussian distributions with mean 
$[140,20,36]$ and standard deviation $[10,10,10]$.
For each realization, we compute the converged estimates at time $t=400$ s and report 
statistics over $N_s=50$ runs. Table~\ref{tab:all_filters_mc_results} summarizes the ensemble 
mean and standard deviation of the final estimation errors in 
$J_x$, $J_y$, and $J_z$, expressed as percentages of their nominal 
values, namely
\begin{equation}
\frac{1}{N_s}\sum_{k=1}^{N_s} 
\left(
\frac{\mathbb{E}\left\{J^{\text{est}}_{\mathcal{C},k}\right\}}
{J_{\mathcal{C}}^\star} - 1
\right)\times 100,
\label{Eq20}
\end{equation}
where $\mathbb{E}\left\{J^{\text{est}}_{\mathcal{C},k}\right\}$ denotes 
the estimated inertia at the final 
time $t=400$ s, for the $k$-th realization, and $J_{\mathcal{C}}^\star$ is 
the corresponding nominal inertia defined in \eqref{eq:Jtrue}.
It is seen that under full excitation, EKF and UKF achieve accurate estimates ($\approx 1\%$ errors), while EnKF exhibits larger errors and variability. The Boosted UKF performs best, with sub-$1\%$ errors and improved consistency due to the WFM-informed update. In the windowed excitation case, the performance of standard filters degrade significantly, particularly EKF/UKF and even more so EnKF, reflecting the limited information available. In contrast, the Boosted UKF maintains high accuracy, indicating that the learned prior effectively regularizes the problem. Finally, under persistent excitation, standard filters still show sensitivity to random initialization, whereas the Boosted UKF consistently achieves the lowest errors and variability.

Overall, these results show that the Boosted UKF is the most robust across all regimes, especially under sparse or intermittent excitation, while EKF/UKF remain competitive only under full excitation.

\section{Summary}
\label{sec:conclusion}
A DT is a virtual counterpart of a physical system that evolves through continuous bidirectional interaction with its physical asset, updating model parameters and associated uncertainties as new measurements are acquired. This feedback loop enables synchronization with the true system dynamics and supports reliable prediction and decision-making. Enhancing this synchronization capability under uncertainty and limited sensing conditions is a central challenge in modern DT architectures.

To address this challenge, we developed a new mathematical framework that integrates WFM  generative modeling with physics-informed nonlinear filtering to improve parameter estimation and uncertainty quantification. The proposed algorithm introduces a confidence-adaptive WFM formulation that dynamically reweights training samples, guiding the generative model toward parameter regimes most informative of the evolving system state.
Our main contributions are threefold: (i) the development of a confidence-adaptive WFM framework capable of dynamic sample reweighting during generative training; (ii) the integration of this learned probabilistic structure into a physically regularized nonlinear filtering architecture, which we called the Boosted UKF; and (iii) the demonstration of superior performance of the Boosted UKF in estimating spacecraft principal moments of inertia under multiple excitation regimes and noise conditions.
The proposed algorithm improves DT fidelity particularly in excitation--limited regimes, where parameters are weakly identifiable and classical filters may suffer from scale ambiguity or bias. When excitation is sufficiently rich, standard filtering techniques already achieve near-optimal performance, and the additional generative regularization primarily enhances convergence rates rather than steady-state accuracy.

\appendix

\section{I. Unscented Kalman Filter}
\label{sec:UKF-setup}
The Unscented Kalman Filter \cite{wan2000unscented} is a derivative-free method for nonlinear state estimation that approximates the propagation of mean and covariance through nonlinear dynamics using a deterministic set of sigma points. In contrast to linearization-based methods, the UKF captures higher-order moments of the underlying distribution, leading to improved accuracy in nonlinear regimes. In this appendix, we outline the UKF formulation used in this work and describe its implementation based on the scaled unscented transform.

The UKF  is initialized at $k=0$ with $(\hat{x}_0^{+},P_0^{+})$, and the recursion below is applied for $k=1,2,\ldots$. For $k\ge 1$, $(\hat{x}^{+}_{k-1}, P^{+}_{k-1})$ denotes the posterior mean and covariance after the $(k-1)$th correction step, which serve as the prior for the current prediction step.
From this prior, we generate a set of scaled sigma points $\{X^{(i)}_{k-1}\}_{i=0}^{2n}$ using the Van der Merwe scaled unscented transform, parameterized by $\alpha$, $\beta$, and $\kappa$, with scaling factor $\lambda=\alpha^2(n+\kappa)-n$, where $n$ denotes the state dimension. The parameters $\alpha$, $\beta$, and $\kappa$ control the spread and weighting of the sigma points; specifically, $\alpha$ determines their dispersion about the mean, $\beta$ encodes prior distribution knowledge, and $\kappa$ is a secondary scaling factor. Since our state vector $x_k$ includes both the physical state variables and the unknown model parameters to be estimated, the sigma points are constructed in the augmented state space. The central and symmetric sigma points are
\begin{equation}
\label{eq:sigma-point-0}
X_{k-1}^{(0)} = \hat{x}_{k-1}^+,
\end{equation}
\begin{equation}
\label{eq:sigma-point-i}
X_{k-1}^{(i)} = \hat{x}^+_{k-1} + s_i,
\qquad
X_{k-1}^{(i+n)} = \hat{x}^+_{k-1} - s_i,
\quad i = 1, \ldots, n,
\end{equation}
where, in \eqref{eq:sigma-point-0} and \eqref{eq:sigma-point-i}, $s_i = S_{k-1} e_i$, $S_{k-1}$ is the Cholesky factor of
$(n+\lambda)P^+_{k-1}$ and $e_i$ denotes the $i$th unit 
vector in $\mathbb{R}^n$. 
The superscript ``$(0)$'' denotes the central sigma point, 
while $(i)$ and $(i+n)$
index the symmetric pairs about the mean, ensuring that the sigma-point set
captures the mean and covariance of the prior distribution and achieves
third-order accuracy for Gaussian inputs under nonlinear mappings. For the first prediction step (from $k=0$ to $k=1$), the sigma points are generated from an initial Gaussian prior $(\hat{x}_0^{+}, P_0^{+})$, which specifies the probabilistic initial condition of the augmented state. Prior knowledge is used to set both $\hat{x}_0^{+}$ and $P_0^{+}$: the physical-state components of $\hat{x}_0^{+}$ are chosen from the assumed initial condition of the physical system, and the parameter components are set to a nominal (design-informed) initial guess; the covariance $P_0^{+}$ encodes the corresponding uncertainty in these initial conditions. 

If a sensor reading at the initial time $k=0$ is available, it is used only to set the physical-state components of $\hat{x}_0^{+}$ (as an initialization, not as a measurement update at $k=0$). Each sigma point is then propagated through the discrete process model to obtain the predicted sigma points, namely,
\begin{equation}
X^{(i)}_{k|k-1} = f\left(X^{(i)}_{k-1}, \tau_k, \Delta t\right),
\qquad i = 0,1,\ldots,2n.
\end{equation}
Here, $(k|k-1)$ denotes the prediction from time $k-1$ to $k$ prior to the
measurement update. Consistent with the discussion after equation \eqref{U1}, the mapping $f(\cdot)$ advances each sigma point one time step, from $k-1$ to $k$, using the input $\tau_k$ and step size $\Delta t$. The propagated sigma points $\{X^{(i)}_{k|k-1}\}_{i=0}^{2n}$ are then used  to obtain the predicted state mean
\begin{equation}
\label{eq:ukf-mean}
\hat{x}_k^- = \sum_{i=0}^{2n} W_m^{(i)}\, X_{k|k-1}^{(i)},
\end{equation}
 and covariance
\begin{equation}
\label{eq:ukf-covariance}
P_k^- = \sum_{i=0}^{2n} W_c^{(i)}
\big(X_{k|k-1}^{(i)} - \hat{x}_k^-\big)
\big(X_{k|k-1}^{(i)} - \hat{x}_k^-\big)^\top
+ Q_k,
\end{equation}
where $W_m^{(i)}$ and $W_c^{(i)}$ are the standard unscented transform weights and $Q_k$ is the process noise covariance (see \eqref{U1}).  Following a similar procedure for the measurement equation \eqref{M1}, we obtain the sigma points $Z_k^{(i)}=h\left(X_{k|k-1}^{(i)}\right)$. Their weighted mean gives the predicted measurement 
\begin{equation}
\hat{z}_k = \sum_{i=0}^{2n} W_m^{(i)} Z_k^{(i)}.
\end{equation}
The innovation covariance and cross covariance are
\begin{equation}
\label{eq:ukf-innovation}
S_k = \sum_{i=0}^{2n} W_c^{(i)}
\big(Z_k^{(i)} - \hat{z}_k\big)
\big(Z_k^{(i)} - \hat{z}_k\big)^\top
+ R_k,
\end{equation}
\begin{equation}
\label{eq:ukf-crosscov}
P_{xz,k} = \sum_{i=0}^{2n} W_c^{(i)}
\left(X^{(i)}_{k|k-1} - \hat{x}_k^-\right)
\left(Z_k^{(i)} - \hat{z}_k\right)^\top.
\end{equation}
Finally, the Kalman gain and sensor correction are
\begin{equation}
\label{eq:ukf-gain}
K_k = P_{xz,k}\, S_k^{-1},
\end{equation}
\begin{equation}
\hat{x}_k^{(1)} = \hat{x}_k^- + K_k\,(z_k - \hat{z}_k), \qquad
P_k^{(1)} = P_k^- - K_k\,S_k\,K_k^\top,
\label{eq:update11}
\end{equation}
where the superscript ``$(1)$'' denotes the posterior after 
the sensor update.

\section{II. Flow Matching and its variants}
\label{sec:appendix1}
In this appendix, we briefly review the FM algorithm \cite{lipman2022flow} and introduce a weighted variant designed to integrate generative modeling with UKF-based state and parameter estimation.
To this end, let $q:\mathbb{R}^d \to \mathbb{R}_{\ge 0}$ be an unknown target probability density from which samples are available. We seek a time-dependent probability density
function  \[
p : [0,1]\times \mathbb{R}^d \to \mathbb{R}_{\ge 0}, 
\qquad (s,x)\mapsto p(s,x),
\]
with the following properties
\begin{equation}
p(0,x) = p_0(x), 
\qquad 
p(1,x) \approx q(x),\quad \forall x\in \mathbb{R}^d, 
\end{equation}
where $p_0:\mathbb{R}^d \to \mathbb{R}_{\ge 0}$ is a prescribed reference density. 
In other words, we seek a transformation that maps a prescribed reference density $p_0$, typically chosen as $p_0(x)=\mathcal{N}(0,I_d)$, to the unknown target density $q$, using only samples drawn from $q$.
To achieve this goal within the flow matching framework, we represent the transformation in terms of the flow generated by the parametric ordinary differential equation
\begin{equation}
\frac{dx}{ds} = v\bigl(s,x;\Theta\bigr), 
\qquad 
x(0)=x_0,
\label{ODEFM}
\end{equation}
where $x:[0,1]\to \mathbb{R}^d$, $v : [0,1]\times \mathbb{R}^d \times \mathbb{R}^b \to \mathbb{R}^d$ is the vector field, and $\Theta \in \mathbb{R}^b$ is the parameter vector.
The vector field $v(s,x;\Theta)$ is assumed to be continuous in $s\in [0,1]$, Lipschitz continuous with respect to $x\in {\mathbb{R}^d}$, and continuously differentiable with respect to $\Theta\in \mathbb{R}^b$. 
In practice, $v$ is represented by a multi-layer perceptron, which provides a flexible representation satisfying all regularity assumptions.

As is well known, if $x_0$ in \eqref{ODEFM} is distributed according to $p_0$, then $x(s)$ admits a probability density $p(s,x)$ for each $s \in [0,1]$. In fact, the initial density $p_0$ is transported to $p(s,x)$ by the flow $\phi(s,x_0)$ defined by \eqref{ODEFM}, which is at least continuous in $x_0$.
If $v(s,x;\Theta)$ is continuously differentiable with respect to $x$, then the flow $\phi(s,x_0)$ is also continuously differentiable with respect to $x_0$. In this case, the density $p(s,x)$ satisfies the Liouville (continuity) equation.
\begin{equation}
\frac{\partial p}{\partial s}(s,x)
+
\text{div}\left[ p(s,x)\, v(s,x;\Theta) \right]=0. 
\label{eq:continuity}
\end{equation}
Here $\text{div}(\cdot)$ denotes the divergence operator for vector fields in $\mathbb{R}^d$. We seek to identify $v$ such that $p$ approximates $q$ at $s=1$.
To compute $v$ based on $M_0$ independent samples of $p_0$ and $M_1$ independent samples of $q$, i.e., 
\begin{align}
x_0^{i} &\sim p_0, \qquad i=1,\ldots,M_0,\\
x_1^{j} &\sim q, \,\,\qquad j=1,\ldots,M_1,
\end{align}
we define, for $s \in [0,1]$, an interpolation between $x_0^{i}$ and $x_1^{j}$ by
\begin{equation}
x_{ij}(s) = (1 - (1-\epsilon_{\min})\,s)\,x_0^{i} \;+\; s\,x_1^{j},
\label{eq:interp}
\end{equation}
where $\epsilon_{\min} > 0$ is a small regularization parameter. This path 
corresponds to the so-called optimal transport (OT) interpolation 
proposed in \cite{lipman2022flow}.
The associated Eulerian target velocity 
along this path is defined as
\begin{equation}
u(s,x_0^{i},x_1^{j})
= \frac{x_1^{j} - (1-\epsilon_{\min})\,x_0^{i}}{1 - (1-\epsilon_{\min})\,s}.
\label{eq:target_velocity}
\end{equation}
The parameters $\Theta$ defining the vector 
field $v$ are obtained by solving the 
non-convex optimization problem
\begin{equation}
\min_{\Theta} \;
\sum_{i=1}^{M_0}\sum_{j=1}^{M_1}
\sum_{k=1}^{M_2}
\left\|
v\bigl(s_k, x_{ij}(s_k);\Theta\bigr)
-
u\bigl(s_k, x_0^{i}, x_1^{j}\bigr)
\right\|^2,
\label{eq:fm_loss_empirical}
\end{equation}
where $\{s_k\}_{k=1}^{M_2}$ are independent samples drawn from the uniform distribution on $[0,1]$.
With $v$ at hand, samples from the target distribution $q$ can be generated by integrating the ODE \eqref{ODEFM} numerically up to $s=1$ for initial conditions $x_0$ drawn from $p_0$.

\subsection{Weighted Flow Matching (WFM)}
\label{sec:weighted-fm}

In many engineering applications, samples from a probability 
distribution may carry different levels of information, reliability, or physical 
consistency. This motivates a WFM formulation in which each sample 
is assigned a nonnegative weight reflecting its relative importance. 
Let $\{x_1^{j}\}_{j=1}^{M_1}$ be samples drawn from $q$, and let $\{w_j\}_{j=1}^{M_1}$ be associated weights satisfying $w_j \ge 0$ and
\[
\sum_{j=1}^{M_1} w_j = 1.
\]
The weights $w_j$ modulate the contribution of each sample $x_1^{j}$ to the regression problem, and will be computed with the algorithm described in Appendix II-\ref{sec:non-linear-weights}.
Equivalently, training samples can be drawn with probabilities proportional to $w_j$, thereby emphasizing more reliable or informative samples, particularly in high-dimensional or noisy settings.
The weighted formulation modifies the regression problem by assigning a weight $w_j$ to each sample $x_1^{j}$. The resulting optimization problem becomes
\begin{equation}
\min_{\Theta} \;
\sum_{i=1}^{M_0}\sum_{j=1}^{M_1}
\sum_{k=1}^{M_2}
w_j\,
\left\|
v\bigl(s_k, x_{ij}(s_k);\Theta\bigr)
-
u\bigl(s_k, x_0^{i}, x_1^{j}\bigr)
\right\|^2.
\label{eq:wfm_loss}
\end{equation}
Equivalently, the weights $w_j$ can be interpreted as defining a non-uniform sampling of the data points $x_1^{j}$, with samples drawn proportionally to $w_j$. In this sense, the weighted formulation emphasizes regions of the data distribution associated with larger weights, which may correspond to more reliable or physically consistent samples.
In practice, the weighted loss \eqref{eq:wfm_loss} is approximated using a mini-batch $\mathcal{B} \subset \{1,\ldots,M_0\} \times \{1,\ldots,M_1\} \times \{1,\ldots,M_2\}$ of index triplets $(i,j,k)$. The corresponding optimization problem takes the form
\begin{equation}
\min_{\Theta}\;
\frac{
\displaystyle
\sum_{(i,j,k)\in \mathcal{B}}
w_j\,
\left\|
v\bigl(s_k, x_{ij}(s_k);\Theta\bigr)
-
u\bigl(s_k, x_0^{i}, x_1^{j}\bigr)
\right\|^2
}{
\displaystyle
\sum_{(i,j,k)\in \mathcal{B}} w_j
}.
\label{eq:wfm_minibatch}
\end{equation}
Algorithm~\ref{alg:WeightedFlow} summarizes the WFM training procedure. Relative to standard FM~\cite{lipman2022flow}, the weighted formulation incorporates per-sample weights to emphasize reliable observations.

\begin{algorithm}[t]
\small
\caption{Weighted Flow Matching (WFM)}
\label{alg:WeightedFlow}

\textbf{Initialization:}
\begin{itemize}
    \item \textbf{Data:} Samples $\{x_1^{j}\}_{j=1}^{M_1} \sim q$ with weights $\{w_j\}_{j=1}^{M_1}$.
    \item \textbf{Reference samples:} Draw $\{x_0^{i}\}_{i=1}^{M_0} \sim p_0$.
    \item \textbf{Model:} Initialize parameters $\Theta$ of the vector field $v(s,x;\Theta)$.
    \item \textbf{Hyperparameters:} learning rate $\eta$, number of iterations $S$.
\end{itemize}

\textbf{Training Process:}
\begin{algorithmic}[1]
    \FOR{$n = 1, \dots, S$}

        \STATE \textbf{Mini-batch sampling:}
        Select indices $\mathcal{I}\subset\{1,\dots,M_0\}$, 
        $\mathcal{J}\subset\{1,\dots,M_1\}$, 
        and $\mathcal{K}\subset\{1,\dots,M_2\}$.

        \STATE \textbf{Interpolation and target velocity:}
        For $(i,j,k)\in \mathcal{I}\times\mathcal{J}\times\mathcal{K}$, define
        \[
        x_{ij}(s_k)
        =
        (1-(1-\epsilon_{\min})s_k)x_0^{i}
        + s_k x_1^{j},
        \]
        \[
        u(s_k,x_0^{i},x_1^{j})
        =
        \frac{x_1^{j} - (1-\epsilon_{\min})x_0^{i}}
        {1-(1-\epsilon_{\min})s_k}.
        \]

        \STATE \textbf{Weighted loss:}
        \[
        \mathcal{L}_{\mathrm{WFM}}
        =
        \sum_{i\in\mathcal{I}}
        \sum_{j\in\mathcal{J}}
        \sum_{k\in\mathcal{K}}
        w_j\,
        \left\|
        v(s_k,x_{ij}(s_k);\Theta)
        -
        u(s_k,x_0^{i},x_1^{j})
        \right\|^2.
        \]

        \STATE \textbf{Parameter update:}
        \[
        \Theta \leftarrow \Theta - \eta \,\nabla_{\Theta}\mathcal{L}_{\mathrm{WFM}}.
        \]

    \ENDFOR
\end{algorithmic}
\end{algorithm}

\subsection{Learning to Re-Weight (LRW)}
\label{sec:non-linear-weights}

In Appendix II-\ref{sec:weighted-fm}, we introduced the WFM formulation, where each
sample $x_1^{j}$ of the target distribution $q$ is assigned a scalar weight $w_j \ge 0$.
In the simplest setting, these weights can be prescribed a priori to reflect measurement
reliability or sensor confidence. However, in many applications, the data may exhibit
non-stationary noise or systematic biases that cannot be captured by fixed weights.
To address this issue, we adopt an LRW strategy~\cite{ren2018learning}, in which the
weights $w_j$ are adapted dynamically using feedback from a small, trusted validation
set. The LRW algorithm partitions the data into three subsets: (i) a training set
$\mathcal{D}_{\mathrm{train}}$, possibly corrupted by noise or bias; (ii) a validation
set $\mathcal{D}_{\mathrm{val}}$, assumed to be representative of the true distribution
and used to guide the reweighting; and (iii) a test set $\mathcal{D}_{\mathrm{test}}$,
used only for performance evaluation. Implementation details, including the MLP
architecture, learning rates, and training schedule, are given in Section~\ref{sec:mainLWR}.
For each training sample $n \in \{1,\ldots,N\}$, the input to the MLP classifier is
the scalar $z$-score
\[
  z_{n} = \frac{e_{n}-\mu_{e}}{\sigma_{e}},
\]
where $e_n$ is the discrete mean-squared error defined in~\eqref{errorLRW}, and
$\mu_{e}$, $\sigma_{e}$ are its mean and standard deviation. The binary label
\[
  y_{n} = \chi\!\left[e_{n} > \mathrm{median}\bigl(\{e_{n}\}\bigr)\right],
\]
where $\chi$ is the indicator function, takes the value 1 when the sample’s trajectory error exceeds the median of $\{e_n\}$, and 0 otherwise. We define the per-sample loss as the binary cross-entropy
\begin{align}
  \ell_{n}(\Upsilon)
  =
  -\Bigl[&
    y_{n}\log\sigma\!\bigl(\mathrm{MLP}(z_{n};\Upsilon)\bigr)
    +\nonumber\\
    &  (1-y_{n})\log\!\bigl(1-\sigma\!\bigl(\mathrm{MLP}(z_{n};\Upsilon)\bigr)\bigr)
  \Bigr],
  \label{eq:persample}
\end{align}
where  $\sigma(\cdot)$ is the sigmoid function and $\mathrm{MLP}(z_{n};\Upsilon)$ denotes the output of the MLP classifier with input $z_n$ and parameters $\Upsilon$ 
used in the LRW algorithm.
At each iteration, LRW performs the bilevel optimization step described hereafter. 

For a mini-batch
$\mathcal{B}_{t} \subset \{1,\ldots,N\}$ of size $|\mathcal{B}_{t}|$ 
drawn from $\mathcal{D}_{\mathrm{train}}$, let
${\varepsilon}=[\,\varepsilon_{n}\,]_{n\in\mathcal{B}_{t}}$
denote a vector of auxiliary variables used
to probe each sample's influence on validation performance. 
Define the inner weighted loss
\begin{equation}
  \mathcal{L}_{\mathrm{inner}}(\Upsilon,{\varepsilon})
  =
  \sum_{n\in\mathcal{B}_{t}}
  \varepsilon_{n}\,\ell_{n}(\Upsilon).
  \label{eq:Linner}
\end{equation}
A single differentiable inner update produces the provisional MLP parameters 
\[
  \Upsilon'({\varepsilon})
  =
  \Upsilon
  -
  \eta_{\Upsilon}\,
  \nabla_{\Upsilon}\mathcal{L}_{\mathrm{inner}}(\Upsilon,{\varepsilon}),
\]
where $\eta_{\Upsilon}>0$ is the inner-loop learning rate. Let
\begin{equation}
  \mathcal{L}_{\mathrm{val}}\!\left(\Upsilon'({\varepsilon})\right)
  =
  \frac{1}{|\mathcal{B}_{v}|}
  \sum_{m\in\mathcal{B}_{v}}
  \ell_{m}\!\left(\Upsilon'({\varepsilon})\right)
  \label{eq:Lval}
\end{equation}
denote the unweighted mean loss over a validation mini-batch $\mathcal{B}_{v}$
drawn from $\mathcal{D}_{\mathrm{val}}$, evaluated at the provisional parameters
$\Upsilon'({\varepsilon})$. The contribution of each training sample is
quantified by the meta-gradient
\[
  u_{n}
  =
  -\left.\frac{\partial\,\mathcal{L}_{\mathrm{val}}\!\left(\Upsilon'({\varepsilon})\right)}
        {\partial\,\varepsilon_{n}}\right|_{\varepsilon=0},
\]
from which the nonnegative unnormalized weights are defined by
\[
  \tilde{w}_{n} = \max(u_{n},\,0).
\]
The batch-normalized weights are then
\[
  w_{n}
  =
  \begin{cases}
    \dfrac{\tilde{w}_{n}}
          {\displaystyle\sum_{n'\in\mathcal{B}_{t}}\tilde{w}_{n'}}
    & \text{if }\displaystyle\sum_{n'\in\mathcal{B}_{t}}\tilde{w}_{n'}>0,\\[10pt]
    \dfrac{1}{|\mathcal{B}_{t}|}
    & \text{otherwise,}
  \end{cases}
\]
with the fallback to uniform weights whenever the denominator vanishes. Finally,
$\Upsilon$ is updated by backpropagating through the re-weighted outer loss
\begin{equation}
  \mathcal{L}_{\mathrm{outer}}
  =
  \sum_{n\in\mathcal{B}_{t}}
  w_{n}\,\ell_{n}(\Upsilon), 
  \label{eq:Louter}
\end{equation}
i.e., 
\begin{equation}
\Upsilon \leftarrow \Upsilon - \eta_{\Upsilon}\,\nabla_{\Upsilon}\mathcal{L}_{\mathrm{outer}}(\Upsilon).
\end{equation}
The full procedure is summarized in Algorithm~\ref{alg:LRW}.
\begin{algorithm}[t]
\small
\caption{Learning to Re-Weight (LRW)}
\label{alg:LRW}
\textbf{Initialization:} Initialize MLP parameters $\Upsilon$.
\begin{algorithmic}[1]
  \FOR{each mini-batch $\mathcal{B}_{t}\subset\mathcal{D}_{\mathrm{train}}$}

    \STATE \textbf{Inner step.}
    Form the weighted loss
    \[
      \mathcal{L}_{\mathrm{inner}}(\Upsilon,{\varepsilon})
      =
      \sum_{n\in\mathcal{B}_{t}}
      \varepsilon_{n}\,\ell_{n}(\Upsilon)
    \]
    and perform one differentiable gradient step to obtain provisional
    parameters
    \[
      \Upsilon'({\varepsilon})
      =
      \Upsilon
      -
      \eta_{\Upsilon}\,\nabla_{\Upsilon}\mathcal{L}_{\mathrm{inner}}(\Upsilon,{\varepsilon}).
    \]

    \STATE \textbf{Meta-gradient step.}
    Draw a validation mini-batch $\mathcal{B}_{v}$ from $\mathcal{D}_{\mathrm{val}}$
    and compute the unweighted mean validation loss
    \[
      \mathcal{L}_{\mathrm{val}}\!\left(\Upsilon'({\varepsilon})\right)
      =
      \frac{1}{|\mathcal{B}_{v}|}
      \sum_{m\in\mathcal{B}_{v}}
      \ell_{m}\!\left(\Upsilon'({\varepsilon})\right).
    \]
    Compute meta-gradients
    \[
      u_{n}=-\left.\frac{\partial\mathcal{L}_{\mathrm{val}}}{\partial\varepsilon_{n}}\right|_{\varepsilon=0}
    \]
    for all $n\in\mathcal{B}_{t}$.

    \STATE \textbf{Weight update.}
    Set $\tilde{w}_{n}=\max(u_{n},0)$ and normalize:
    \[
      w_{n}
      =
      \begin{cases}
        \tilde{w}_{n}\Big/{\displaystyle\sum_{n'\in\mathcal{B}_{t}}\tilde{w}_{n'}}
        & \text{if the sum}>0,\\[6pt]
        1/|\mathcal{B}_{t}| & \text{otherwise.}
      \end{cases}
    \]

    \STATE \textbf{Outer step.}
    Form the re-weighted loss
    \[
      \mathcal{L}_{\mathrm{outer}}
      =
      \sum_{n\in\mathcal{B}_{t}}
      w_{n}\,\ell_{n}(\Upsilon)
    \]
    and update $\Upsilon$ via
    \[
      \Upsilon \leftarrow \Upsilon - \eta_{\Upsilon}\,\nabla_{\Upsilon}\mathcal{L}_{\mathrm{outer}}(\Upsilon).
    \]

  \ENDFOR
\end{algorithmic}
\end{algorithm}
Each iteration involves only a single differentiable inner update, avoiding the
computational burden of fully unrolled second-order optimization. This adaptive
formulation aligns the weighting of training samples with validation performance,
thereby improving robustness to noise and model bias.

\bibliography{references}

\begin{IEEEbiography}[{\includegraphics[width=1in,height=1.25in,clip,keepaspectratio]{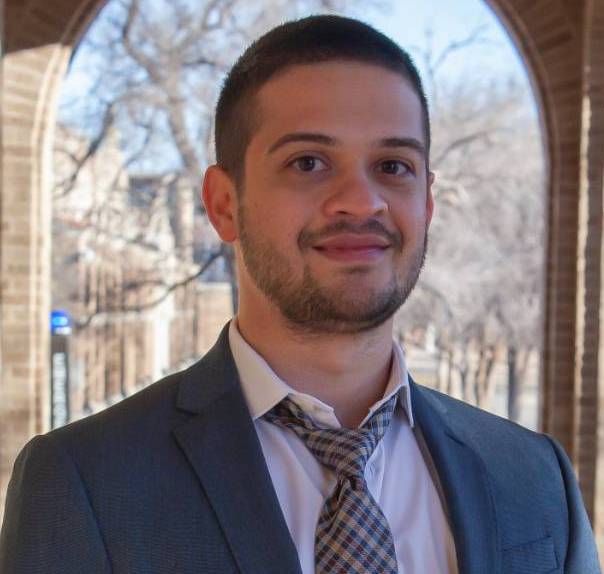}}]{YASAR YANIK} received the B.S. degree in mechanical engineering from Istanbul Technical University, Istanbul, Turkey, in 2016, the M.S. degree in mechanical engineering from São Paulo State University (Universidade Estadual Paulista), São Paulo, Brazil, in 2019, and the Ph.D. degree in mechanical engineering from Texas Tech University, Lubbock, TX, USA, in 2024.
He is currently a Postdoctoral Researcher in the Department of Applied Mathematics at the University of California, Santa Cruz, where he leads research on Digital Twin Enabled Autonomous Control for On-Orbit Spacecraft Servicing (SURI), supported by the U.S. Department of Defense and Raytheon. His primary research area is digital twin modeling, focusing on the integration of physics-based simulation, machine learning, and uncertainty quantification for autonomous systems.
Dr. Yanik’s work spans rotating machinery, photovoltaic systems, and spacecraft subsystems, developing digital twin frameworks that enable predictive maintenance, performance optimization, and health monitoring through physics-informed AI and generative modeling. He has collaborated with Pantex, Sandia National Laboratories, and the Air Force Research Laboratory to advance data-driven modeling, distributed control, and system identification. His research contributions have been published in leading journals, and his expertise in machine-learning-driven forecasting, anomaly detection, and digital twin verification provides practical insights for engineering applications. Dr. Yanik is also a member of the ASME and Phi Kappa Phi Honor Society.
\end{IEEEbiography}
\vspace{-8ex}
\begin{IEEEbiography}[{\includegraphics[width=1in,height=1.25in,clip,keepaspectratio]{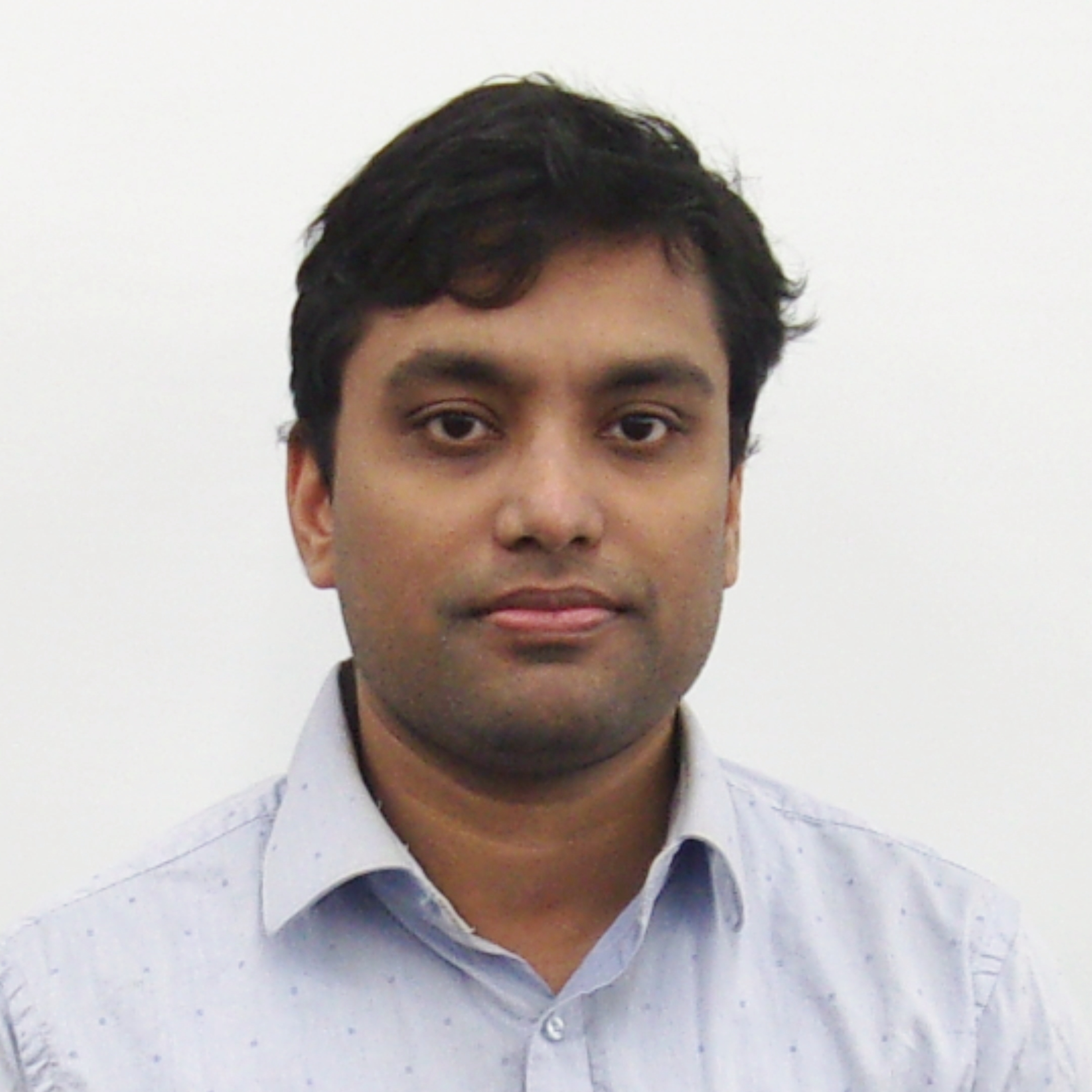}}]{Himadri Basu} received his PhD degree in Electrical and Computer Engineering, Durham, New Hampshire in September 2020. He is currently a Postdoctoral Researcher with the Hybrid Systems Laboratory (HSL), University of California, Santa Cruz, CA, USA. His research interests include distributed control of multi-agent systems, cooperative guidance, estimation and control, with applications to spacecraft rendezvous and docking. He has authored and coauthored several papers in these areas. His current work focuses on developing scalable and resilient control architectures for autonomous multi-agent and space systems. Dr. Basu has served as a reviewer for several IEEE journals and conferences in the areas of control and dynamic systems. Dr. Basu is currently leading a research project on Digital Twin–Enabled Autonomous Spacecraft Guidance and Proximity Maneuvering, supported by the Air Force Office of Scientific Research (AFOSR), in collaboration with the Department of Defense and Raytheon.
\end{IEEEbiography}
\vspace{-8ex}
\begin{IEEEbiography}[{\includegraphics[width=1in,height=1.25in,clip,keepaspectratio]{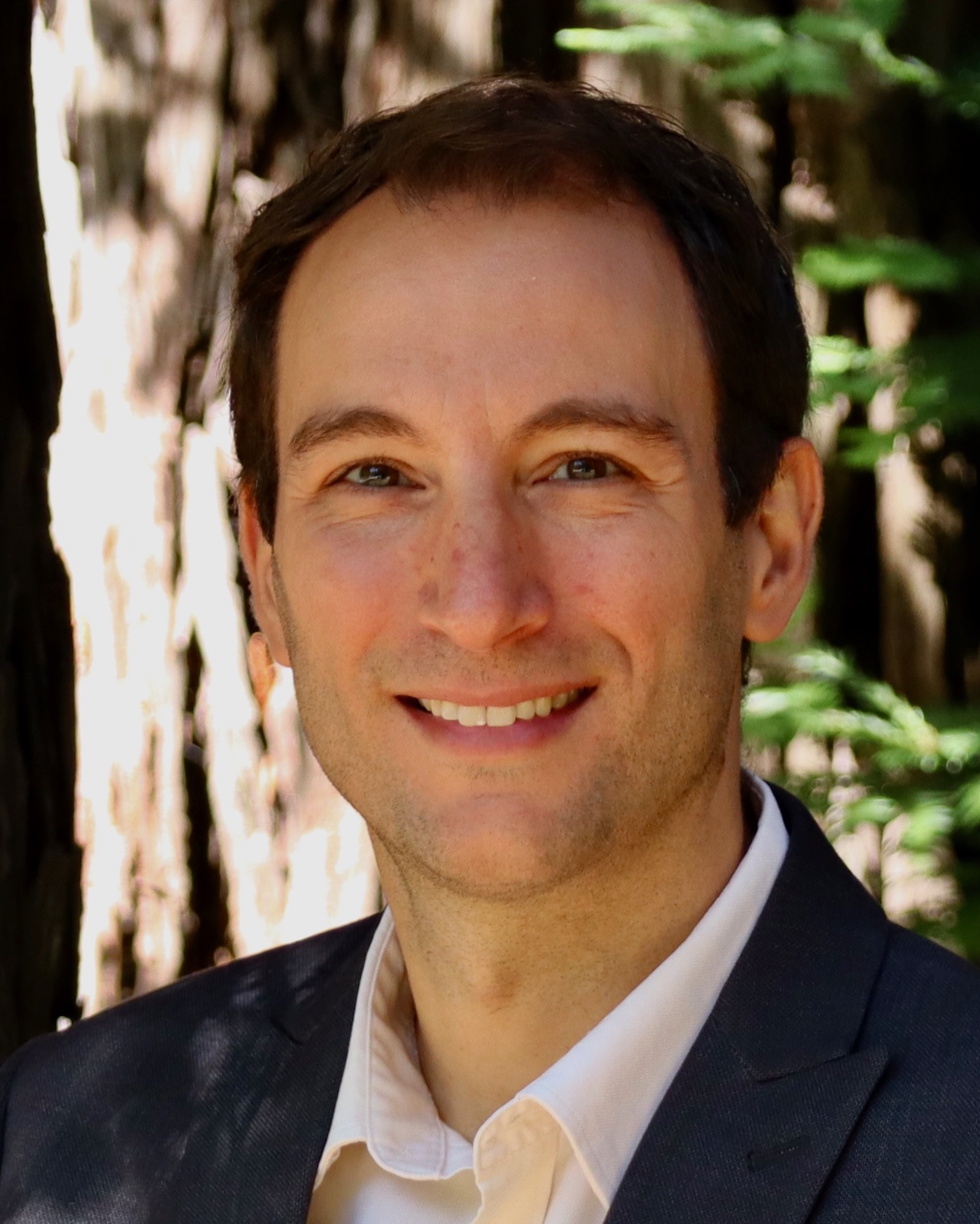}}]{Ricardo Sanfelice} received the B.S. degree in Electronics Engineering from the Universidad de Mar del Plata, Buenos Aires, Argentina, in 2001, and the M.S. and Ph.D. degrees in Electrical and Computer Engineering from the University of California, Santa Barbara, in 2004 and 2007, respectively. In 2007 and 2008, he held postdoctoral positions at the Laboratory for Information and Decision Systems at the Massachusetts Institute of Technology and at the Centre Automatique et Systèmes at the École de Mines de Paris. In 2009, he joined the faculty of the Department of Aerospace and Mechanical Engineering at the University of Arizona, Tucson, where he was an Assistant Professor. In 2014, he joined the University of California, Santa Cruz, where he is currently Professor and Chair in the Department of Electrical and Computer Engineering. Prof. Sanfelice is the recipient of the 2013 SIAM Control and Systems Theory Prize, the National Science Foundation CAREER award, the Air Force Young Investigator Research Award, the 2010 IEEE Control Systems Magazine Outstanding Paper Award, and the 2020 Test-of-Time Award from the Hybrid Systems: Computation and Control Conference. He is Associate Editor for Automatica, Communicating Editor for the Journal of Nonlinear Science, and a Fellow of the IEEE. His research interests are in modeling, stability, robust control, observer design, and simulation of nonlinear and hybrid systems with applications to power systems, aerospace, and biology.
\end{IEEEbiography}
\vspace{-8ex}
\begin{IEEEbiography}[{\includegraphics[width=1in,height=1.25in,clip,keepaspectratio]{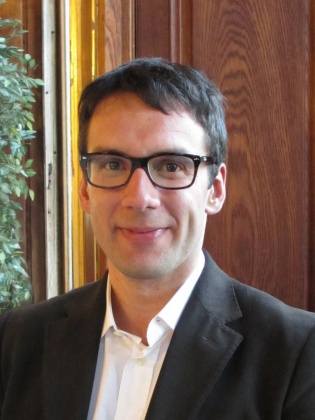}}]{Daniele Venturi}
received the combined B.S. and Sc.M. degrees in Mechanical Engineering from the University of Bologna, Bologna, Italy, in 2002, and the Ph.D. degree in Applied Physics from the same institution in 2006. He was a Postdoctoral Research Associate with the Department of Energy, Nuclear and Environmental Engineering (now Department of Industrial Engineering), University of Bologna, from 2006 to 2010, where he conducted research in theoretical, computational, and experimental fluid dynamics. From 2010 to 2015, he was a Research Assistant Professor of Applied Mathematics at Brown University, Providence, RI, USA. In 2015, he joined the Department of Applied Mathematics, University of California, Santa Cruz, CA, USA, where he is now Full Professor.  His research interests and activities  include stochastic modeling, uncertainty quantification, high-performance scientific computing, mutli-fidelity modeling, and numerical approximation of high-dimensional partial differential equations. 
\end{IEEEbiography}

\EOD
\end{document}